\documentclass[preprint,authoryear]{elsarticle}
\usepackage{graphicx,epsfig,float,rotating,setspace}
\usepackage{natbib,txfonts}
\usepackage{amssymb}
\journal{New Astronomy}

\DeclareRobustCommand{\ion}[2]{%
\relax\ifmmode
\ifx\testbx\f@series
{\mathbf{#1\,\mathsc{#2}}}\else
{\mathrm{#1\,\mathsc{#2}}}\fi
\else\textup{#1\,{\mdseries\textsc{#2}}}%
\fi}
\def\astrobj#1{#1}

\def\aj{AJ}%
\def\apj{ApJ}%
\def\apjs{ApJS}%
\def\aap{A\&A}%
\def\aaps{A\&AS}%
\def\mnras{MNRAS}%
\def\na{New A}%
\def\pra{Phys.~Rev.~A}%
\def\pasp{PASP}%
\def\pasj{PASJ}%
\def\ssr{Space~Sci.~Rev.}%
\def\jqsrt{J.~Quant.~Spec.~Radiat.~Transf.}%
\def\physscr{Phys.~Scr}%
\def\teff{$T_\mathrm{eff}$}
\def\logg{$\log\,g$}

\begin{document}
\begin{frontmatter}

\title{Abundance analysis of the supergiant stars \astrobj{HD\,80057} and \astrobj{HD\,80404} based on their UVES Spectra}
\author[NU]{T. Tanr{\i}verdi\corref{cor1}}
\author[AU]{\"O. Ba\c{s}t\"urk}
\address[NU]{Ni\u{g}de University, Faculty of Arts and Sciences, Department of Physics, TR-51240, Ni\u{g}de, Turkey}
\address[AU]{Ankara University, Faculty of Science, Department of Astronomy and Space Sciences, TR-06100, Tando\u{g}an, Ankara, Turkey}

\cortext[cor1]{e-mail: ttanriverdi@nigde.edu.tr}

\begin{abstract}
\label{abstract}
This study presents elemental abundances of the early A-type supergiant  \astrobj{HD\,80057} and the late A-type supergiant \astrobj{HD\,80404}. High resolution and high signal-to-noise ratio spectra published by the UVES Paranal Observatory Project \citep{bagnulo2003}\footnote{Based on data obtained with the UVES Paranal Observatory Project (ESO DDT Program ID 266.D-5655)} were analysed to compute their elemental abundances using ATLAS9 \citep{kurucz1993, 2005MSAIS...8...14K, sbordone2004}.  In our analysis we assumed local thermodynamic equilibrium. The atmospheric parameters of \astrobj{HD~80057} used in this study are from \citet{FP2012}, and that of \astrobj{HD~80404} are derived from spectral energy distribution, ionization equilibria of \ion{Cr}{i/ii} and \ion{Fe}{i/ii}, and the fits to the wings of  Balmer lines and Paschen lines as \teff\,= 7700\,$\pm$\,150 K and \logg\,=\,1.60\,$\pm$\,0.15 (in cgs). The microturbulent velocities of \astrobj{HD\,80057} and \astrobj{HD\,80404} have been determined as 4.3 $\pm$ 0.1 and 2.2 $\pm$ 0.7 km~s$^{-1}$. The rotational velocities are 15\,$\pm$\,1 and 7\,$\pm$\,2 km~s$^{-1}$ and their macroturbulence velocities are 24\,$\pm$\,2 and 2\,$\pm$\,1 km~s$^{-1}$. We have given the abundances of 27 ions of 20 elements for \astrobj{HD\,80057} and 39 ions of 25 elements for \astrobj{HD\,80404}. The abundances are close to solar values, except for some elements (Na, Sc, Ti, V, Ba, and Sr). We have found the metallicities [M$/$H] for \astrobj{HD\,80057} and \astrobj{HD\,80404} as -0.15 $\pm$\, 0.24 and -0.02 $\pm$ 0.20 dex, respectively. The evolutionary status of these stars are discussed and their nitrogen-to-carbon ($N/C$)  and nitrogen-to-oxygen ($N/O$) ratios show that they are in their blue supergiant phase before the red supergiant region.

\end{abstract}

\end{frontmatter}

\section{Introduction}
\label{introduction}

A-type supergiants are attractive astrophysical targets for chemical abundance studies. First of all, they are among the brightest stars at visual wavelengths, which makes them observable with low exposure times, high resolution and high signal-to-noise ratios (S/N). Moreover, their spectra are unblended, hence the abundances of numerous elements with consecutive ionization levels such as light elements, $\alpha$ process elements, iron group and s-process elements can be derived from studies of their atmospheres \citep{1995ApJS...99..659V, 2000A&A...364..237A, 2002PhDT.........3P, 2008A&A...479..849S, FP2012, tt2013}. With these results in hand, it is possible to understand their nature and the environments in which they exist, and thereby study galactic and extra-galactic abundance gradients and dispersions.

This paper is a continuation of analyses of early and late A-type supergiants started by \citet{2004IAUS..224..869T, tt2013}. In this study, elemental abundance analyses of two A-type supergiants, \astrobj{HD 80057} and \astrobj{HD 80404} (\astrobj{\textit{iota} Car}), and their revised atmospheric parameters, are presented in detail. These analyses are based on spectra distributed by the UVES Paranal Observatory Project \citep{bagnulo2003}.

\subsection{\astrobj{HD\,80057}}
\label{target1/hd80057}

\astrobj{HD\,80057} (\astrobj{HR\,3688}, \astrobj{HIP\,45481}, \astrobj{SAO\,221010}) was classified as A1\,Iab by \citep{FP2012} (see Table \ref{table:parameters} for more information). It is a member of the Vela OB1 association of stars, which is one of the largest OB star associations of the Galaxy, and is composed of numerous members \citep{reed2000}. Although \astrobj{HD\,80057} has been used as a photometric standard star, \citep{menzies89, cousins90}, and as a spectroscopic standard star for its radial velocities \citep{reed97, gontcharov2006}, a detailed study of its atmosphere was published only very recently by \citet{FP2012}. The authors determined the atmospheric parameters (\teff\, and \logg\,) using spectroscopic indicators and spectroscopic data (see Table \ref{table:literature}) in their study, as well as CNO abundances of {HD\,80057} computed using non-LTE methods.

\subsection{\astrobj{HD\,80404}}
\label{target2/hd80404}

\astrobj{\textit{iota} Car} (\astrobj{HD\,80404, HR\,3699}, \astrobj{HIP\,45556}, \astrobj{SAO\,236808}) is an MK Standard, which is classified as A8\,Ib \citep{malaroda1973,monier1999} (see Table \ref{table:parameters} for more information). It is one of the brightest stars in the southern sky in the visual region of the electromagnetic spectrum. \citet{adelman2000IBVS} listed it amongst the least variable Hipparcos targets, while \citet{graygarrison89} gave its Str\"omgren photometric parameters. 

The atmosphere of \astrobj{\textit{iota} Car} was first studied in detail by \citet{boiarchuk1984}, who gave its spectroscopic parameters as 7300\,$\pm$\,200\,K  for its effective temperature and 1.40\,$\pm$\,0.2 for its surface gravity. \citet{luck1985} later gave elemental abundances for the star as well as revising its stellar parameters, finding   \teff = 7500\,$\pm$\,200\,K, \logg = 0.90\,$\pm$\,0.3, and  micro \& macro-turbulent velocities of 2.5, and 1.0\,$\pm$\,0.5km~s$^{-1}$, respectively. Next, \citet{luck1992} adopted the effective temperature found by \citet{luck1985} as 7500\,K, and found a value of 1.6\,$\pm$\,0.2 for its surface gravity using the \texttt{MARCS} code of \citet{gustafsson75} from its Fe I/II ionization balance. Then, \citet{takeda1995} re-calculated the CNO abundances of \astrobj{\textit{iota} Car} using \citet{luck1985}'s atmospheric parameters and equivalent widths (EW). 

\citet{smiljanic2006}, on the other hand, computed the effective temperature to be 7500$\pm$200\,K, surface gravity 2.40$\pm$0.25, and micro-turbulent velocity 2.34$\pm$0.35 km\,s$^{-1}$  based on their high-resolution spectroscopic observations using  fits to the H$\alpha$ wings, the \ion{Fe}{i/ii} ionization equilibrium and their photometric calibration (see Table~\ref{table:literature}). They also determined its C, N, O, and Fe abundances from their FEROS (Fiber-fed Extended Range Optical Spectrograph) spectra in the wavelength range 3500-9200 \AA\, and a resolution, R=48000. Later on, spectrophotometric observations (visual and near-infrared) of the object were presented by \citet{ruban2006}.

\section{The Spectra}
\label{thespectra}
The UVES spectra used in this study were obtained from the UVES-POP database.  They have high resolution (R $\sim$ 80000) and S/N ratios (for most of the spectra S/N ratio $\sim$ 300-500 in V band). They cover a wavelength range of $\lambda \lambda$ 3040-10400 \AA \citep{bagnulo2003}\footnote{http://www.eso.org/sci/observing/tools/uvespop.html}. The ultraviolet (UV), visual and infrared (IR) parts of UVES spectra were used to determine the abundances of elements such as C, N, O, Mg and Al (see Table \ref{table:abundances}). All spectra were continuum normalized using \textit{IRAF}\footnote{IRAF is distributed by the National Optical Astronomy Observatory, which is operated by the Association of Universities for Research in Astronomy (AURA) under cooperative agreement with the National Science Foundation.} task \textit{continuum}. Then, the EWs of the identified lines were measured using the \textit{splot} package within IRAF. Main sources for line identification are mentioned in \citet{tt2013}. International Ultraviolet Explorer (IUE)'s flux-calibrated spectra were downloaded from MAST\footnote{http://archive.stsci.edu} archive. For spectrophotometry of \astrobj{HD\,80404}, low-dispersion and large aperture spectra, \textit{SWP36720} and \textit{LWP15980}, were used. 


\section{Stellar Parameters}
\label{stellarparameters}

\begin{figure*}[ht]
\center
\includegraphics[width=120mm,height=80mm]{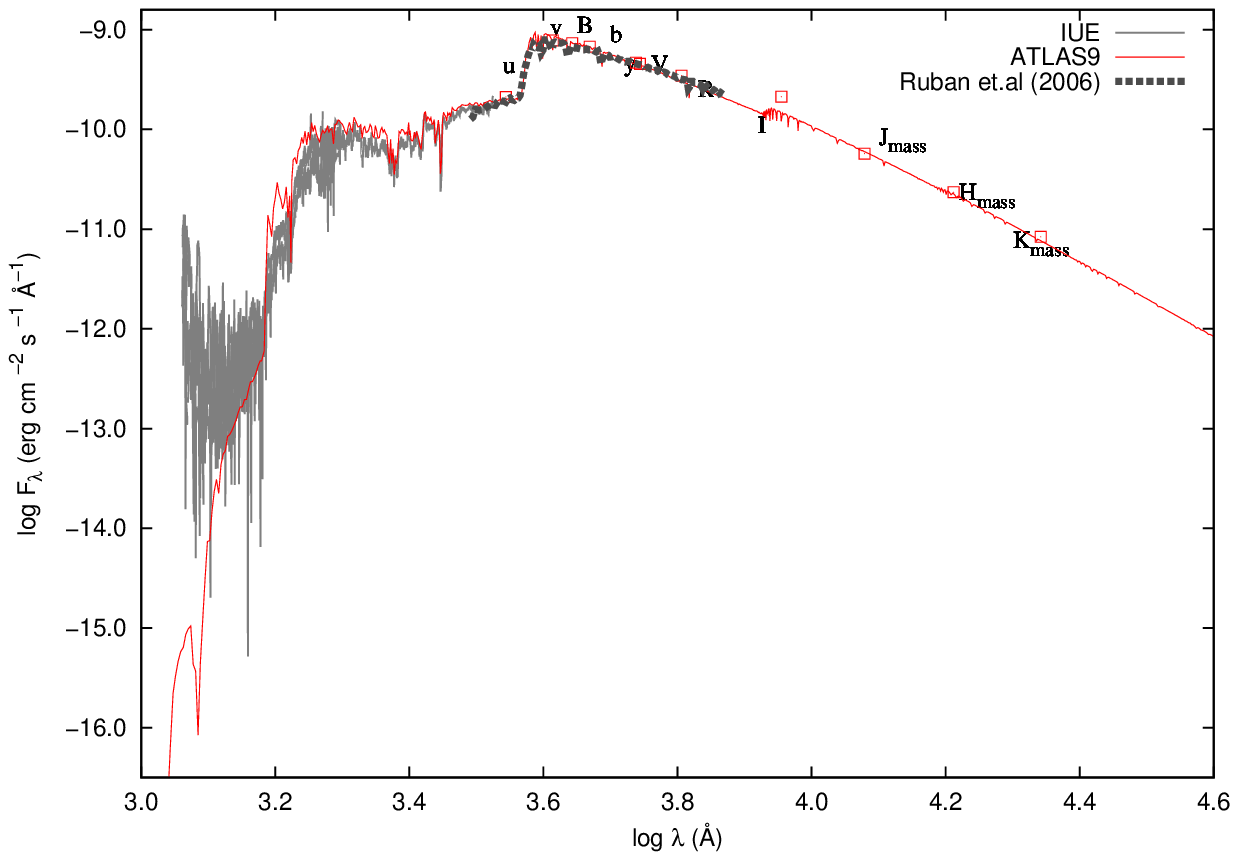}
\caption{A comparison of the observed and computed fluxes (\teff = 7700\,K, \logg = 1.60) for \astrobj{HD\,80404}. ATLAS9 model flux, ATLAS9 reddened model flux, IUE spectra, the spectrophotometric data of \citet{ruban2006} and the photometric data are also given.\label{fig:sed}}
\end{figure*}

\begin{figure}
 \begin{center}
  \includegraphics[width=0.45\textwidth]{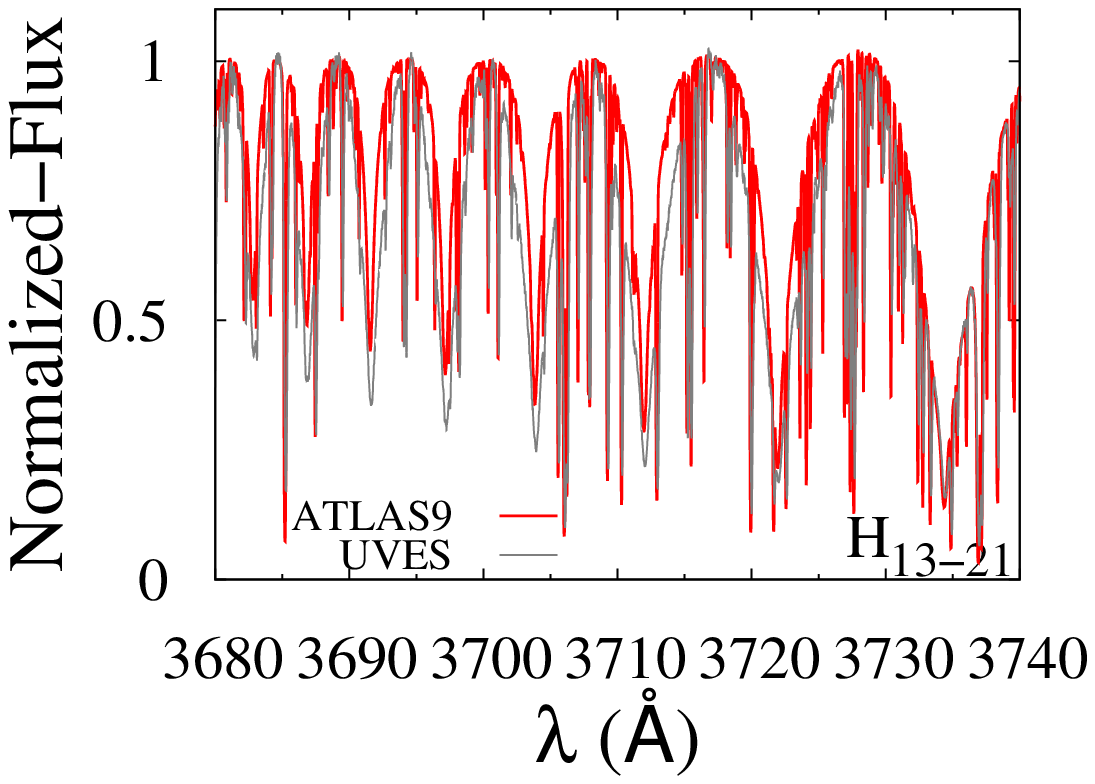}
  \includegraphics[width=0.45\textwidth]{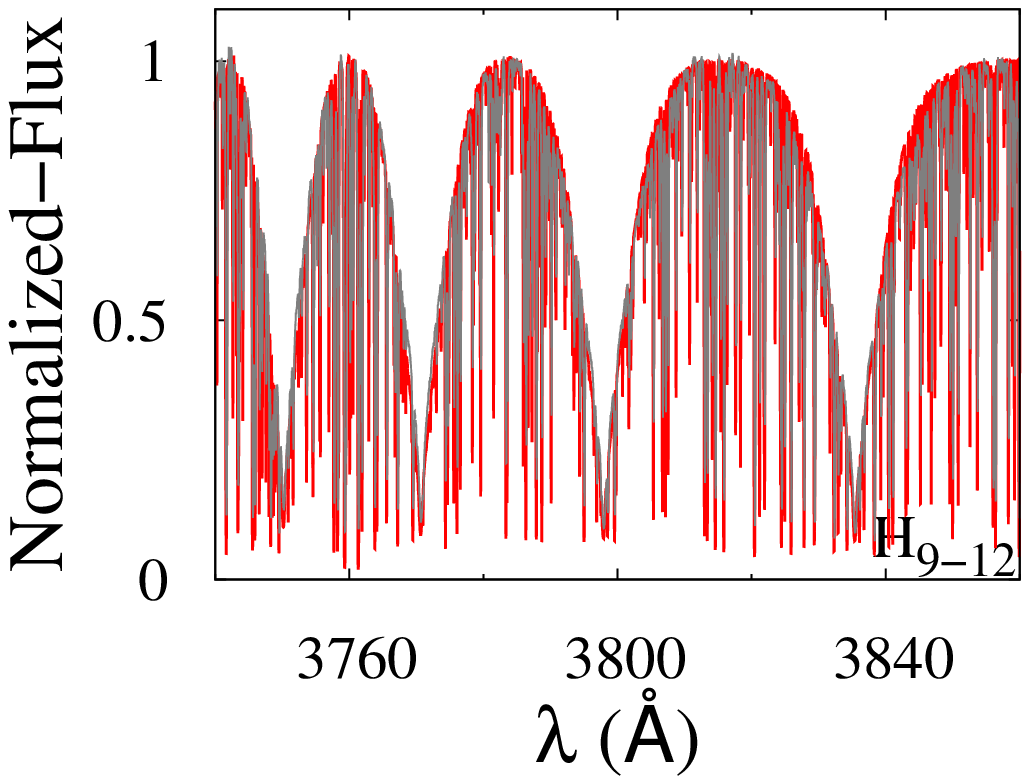}
  \includegraphics[width=0.45\textwidth]{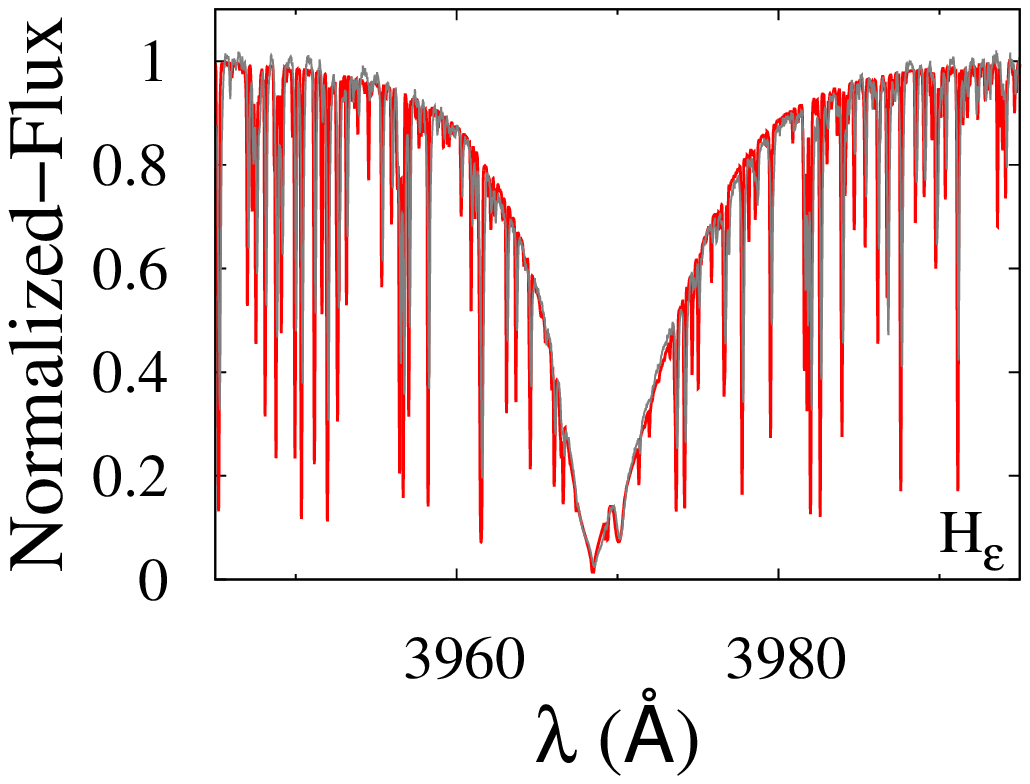}
  \includegraphics[width=0.45\textwidth]{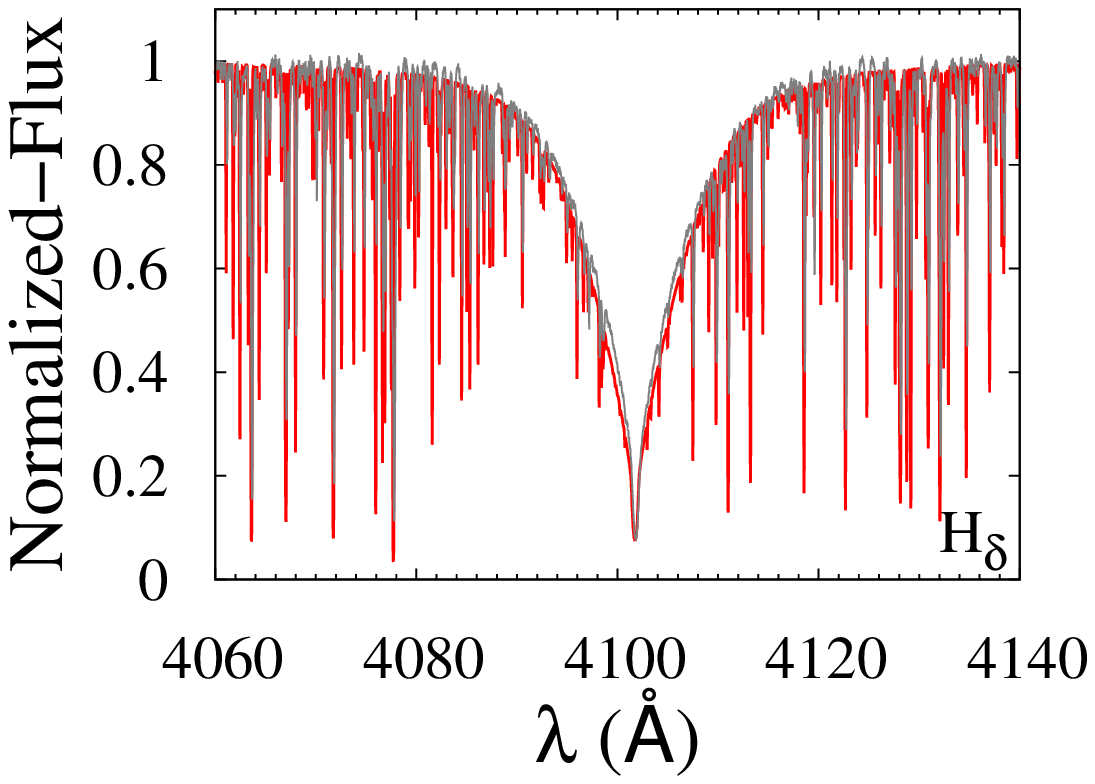}
  \includegraphics[width=0.45\textwidth]{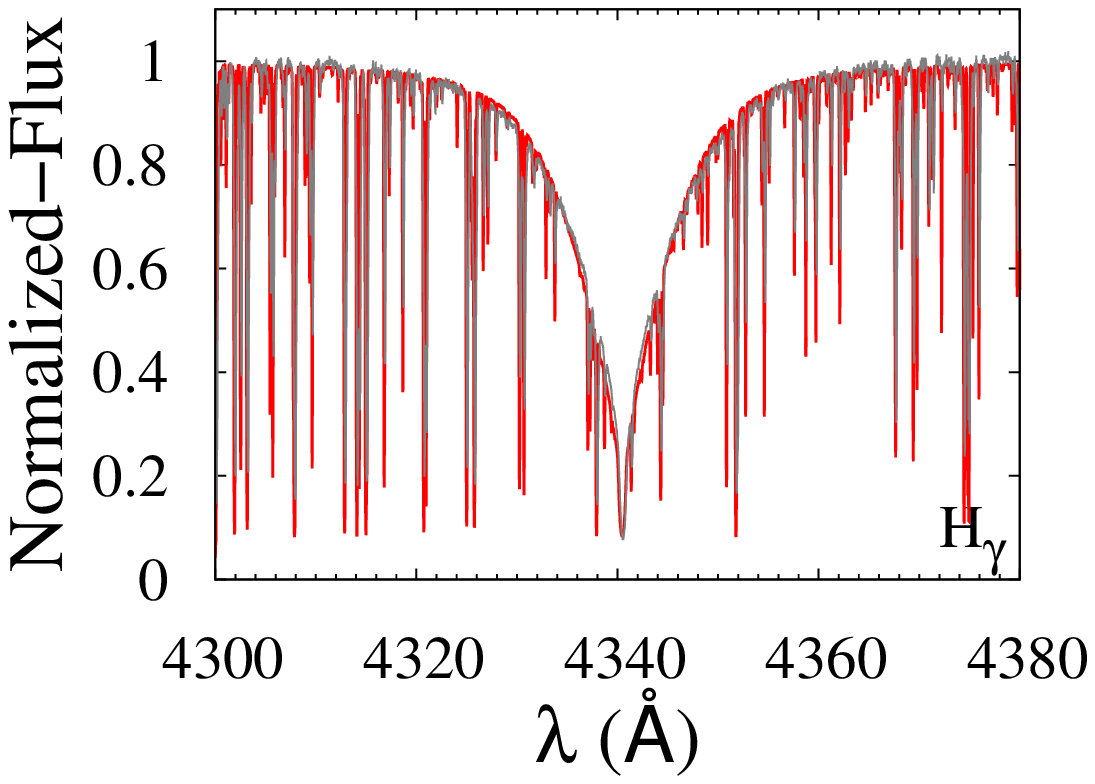}
  \includegraphics[width=0.45\textwidth]{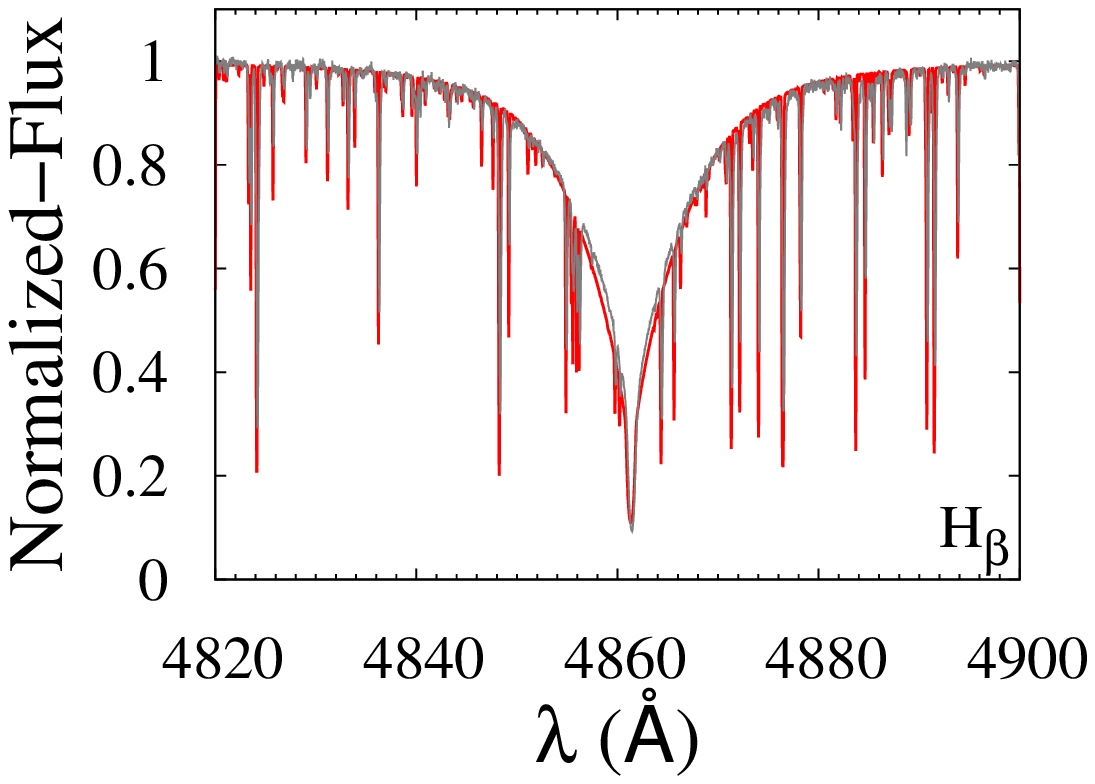}       
  \includegraphics[width=0.45\textwidth]{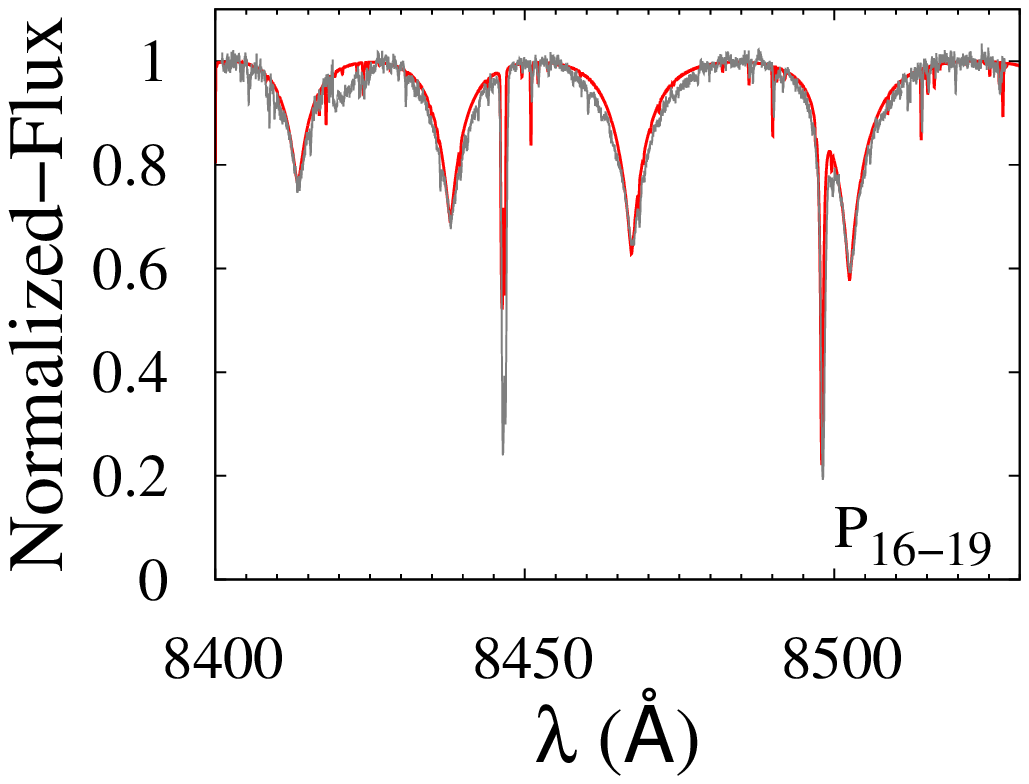}
  \includegraphics[width=0.45\textwidth]{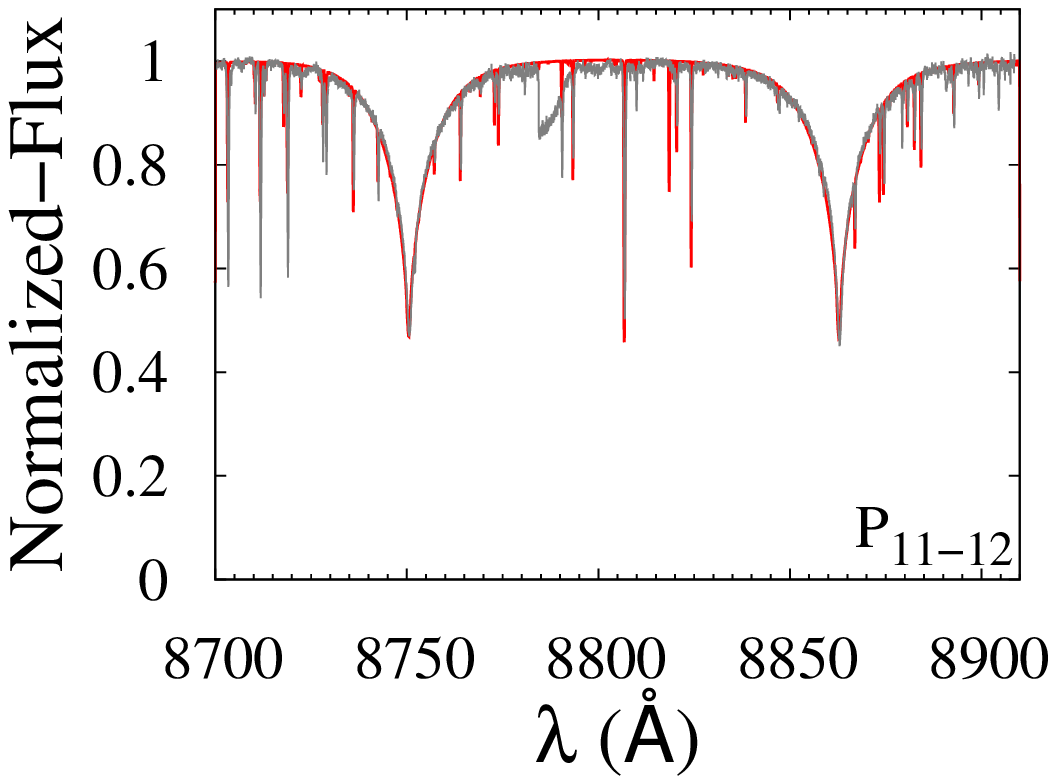}         
 \caption[ ]{Synthetic spectrum fits of Balmer and Paschen series in the UVES spectra of  \astrobj{HD\,80404} for \teff = 7700\,K, \logg = 1.60 pair.\label{fig:paschen}}

 \end{center}
\end{figure}

\begin{figure}
 \begin{center}
  \includegraphics[width=0.45\textwidth]{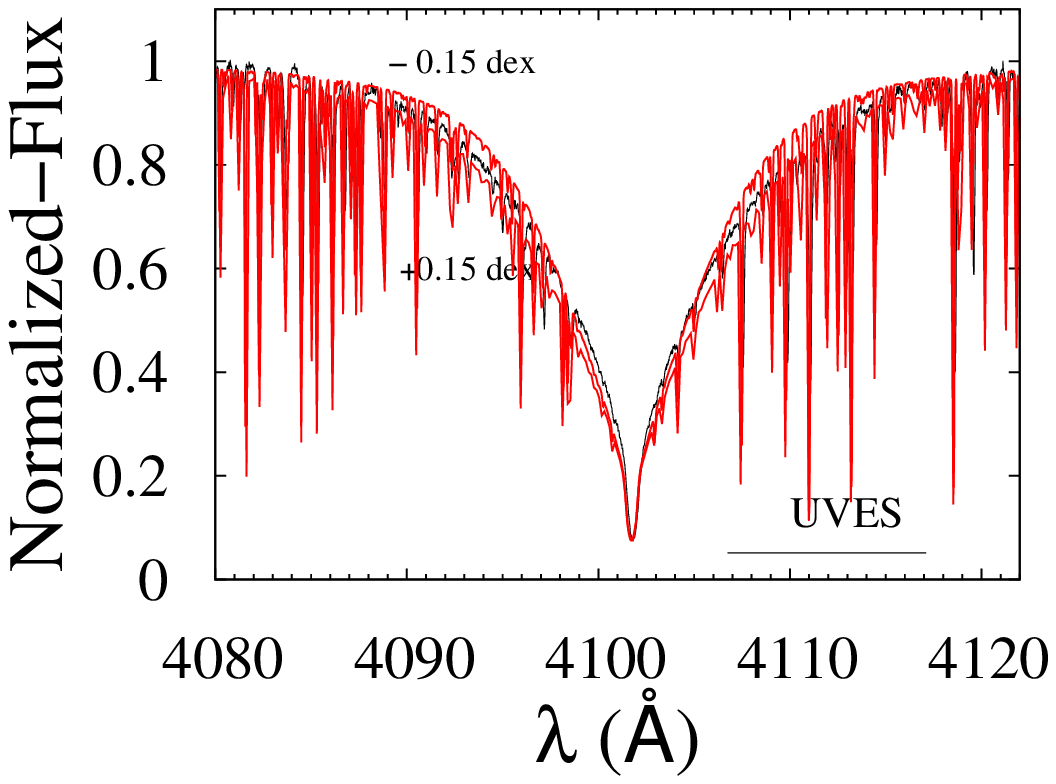}
  \includegraphics[width=0.45\textwidth]{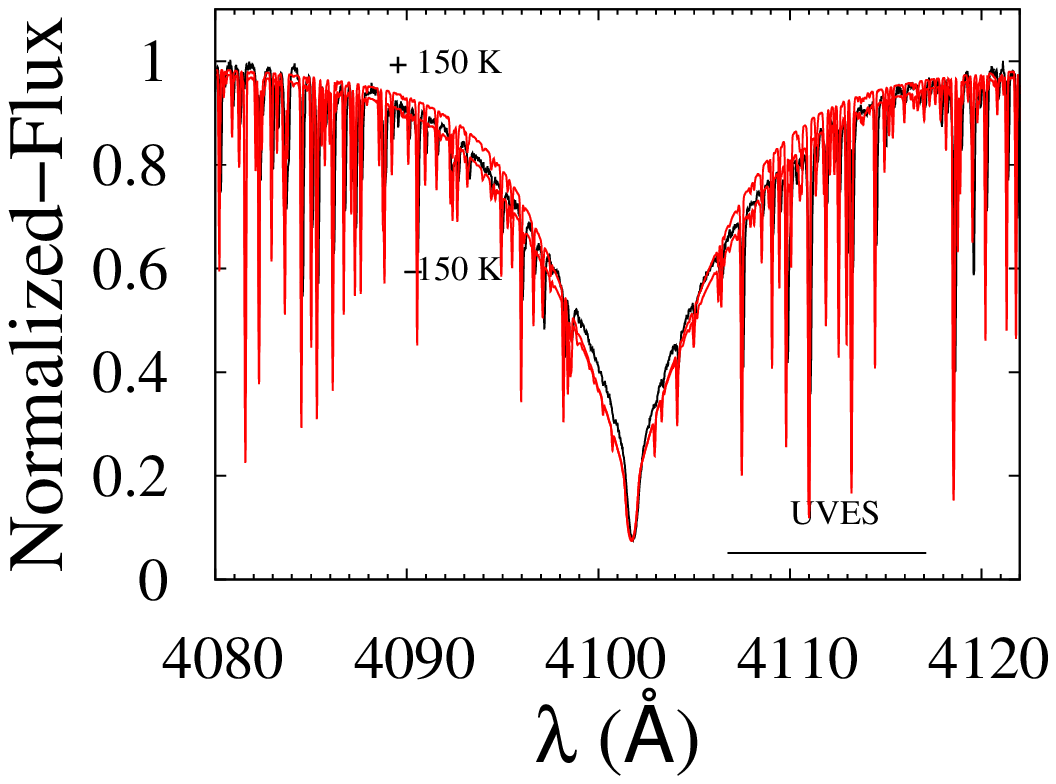}
 \caption[ ]{Synthetic spectrum fits of H$_\delta$ lines in \astrobj{HD\,80404} spectra using \teff = 7700 ($\pm$ 150 K), \logg = 1.60 ($\pm$ 0.15 dex), synthetic spectrum and UVES spectrum.\label{fig:hdelta}}

 \end{center}
\end{figure}

\begin{figure*}

\center
\includegraphics[width=120mm,height=80mm]{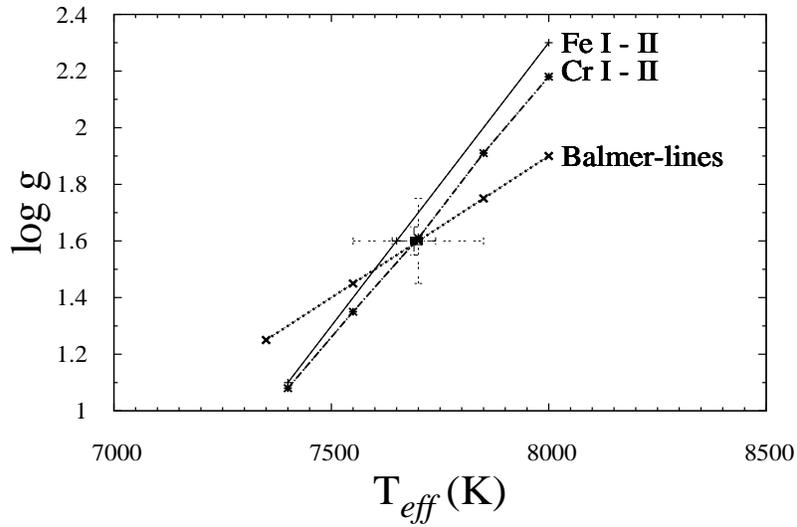}
\caption[]{\teff\, - \logg\, planes for  \teff = 7700\,K, \logg = 1.60 of  \astrobj{HD\,80404} based on \ion{Cr}{i/ii}, \ion{Fe}{i/ii} ionization levels and Balmer line fits.\label{fig:kiel}} 
\end{figure*}

The atmospheric models were produced using \texttt{ATLAS9} \citep{kurucz1993, sbordone2004}. LTE abundance analyses were performed based on EW measurements using the \texttt{WIDTH9} code\citep{kurucz1993}. The effective temperatures and surface gravities (\teff, \logg) were determined via  SED (Spectral Energy Distribution, see Fig.~\ref{fig:sed}) analyses, fits to Balmer and Paschen lines wings (see Figs.~\ref{fig:paschen} \& ~\ref{fig:hdelta}) and from the ionization equilibria of \ion{Cr}{i/ii} and \ion{Fe}{i/ii} (see Fig.~\ref{fig:kiel}). The procedure used to determine these fundamental parameters are illustrated on the \teff -- \logg\ plane in Fig.~\ref{fig:kiel}. The microturbulent velocity was determined by finding the value where the correlation between the derived abundances and the EWs ($\xi_1$) was minimised, and the minimum scatter about the abundance mean ($\xi_2$) was obtained \citep{blackwell1982}. Microturbulent velocities were determined for \astrobj{HD\,80057} and \astrobj{HD\,80404} to be 4.30 and 2.20 km s$^{-1}$, respectively. The derived microturbulent velocities of different species are given in Table~\ref{table:microturbulence}. The rotational velocity and the macroturbulent velocity of \astrobj{HD\,80057} are determined using synthetic spectra produced by \texttt{SYNSPEC} and \texttt{SYNPLOT} \citep{1988CoPhC..52..103H,2011ascl.soft09022H}. The rotational and macroturbulence velocities for \astrobj{HD\,80057} are 15 $\pm$ 1 km~s$^{-1}$ and 24 $\pm$ 2 km~s$^{-1}$  and those for \astrobj{HD\,80404} 
are 7 $\pm$ 2 km~s$^{-1}$ and 2 $\pm$ 1 km~s$^{-1}$.

\begin{figure*}
\centering
\includegraphics[width=12cm,height=5cm]{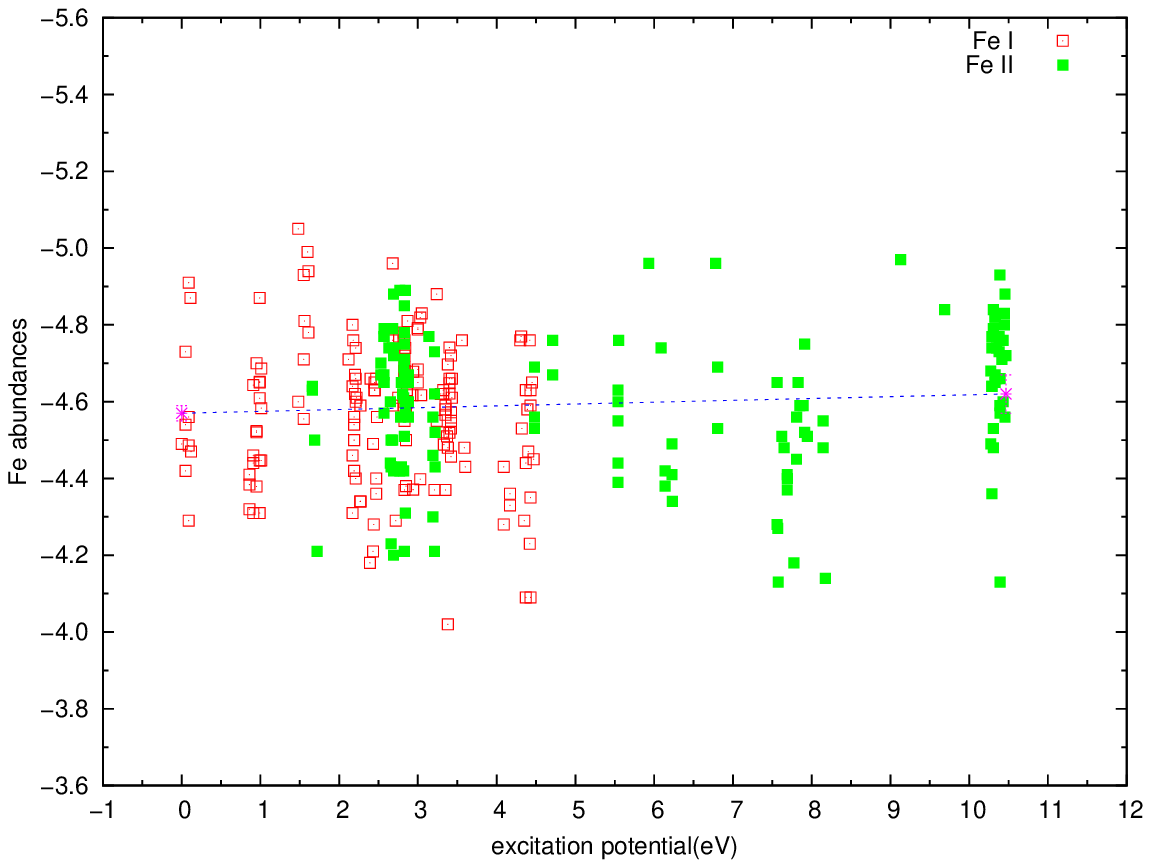}
\includegraphics[width=12cm,height=6cm]{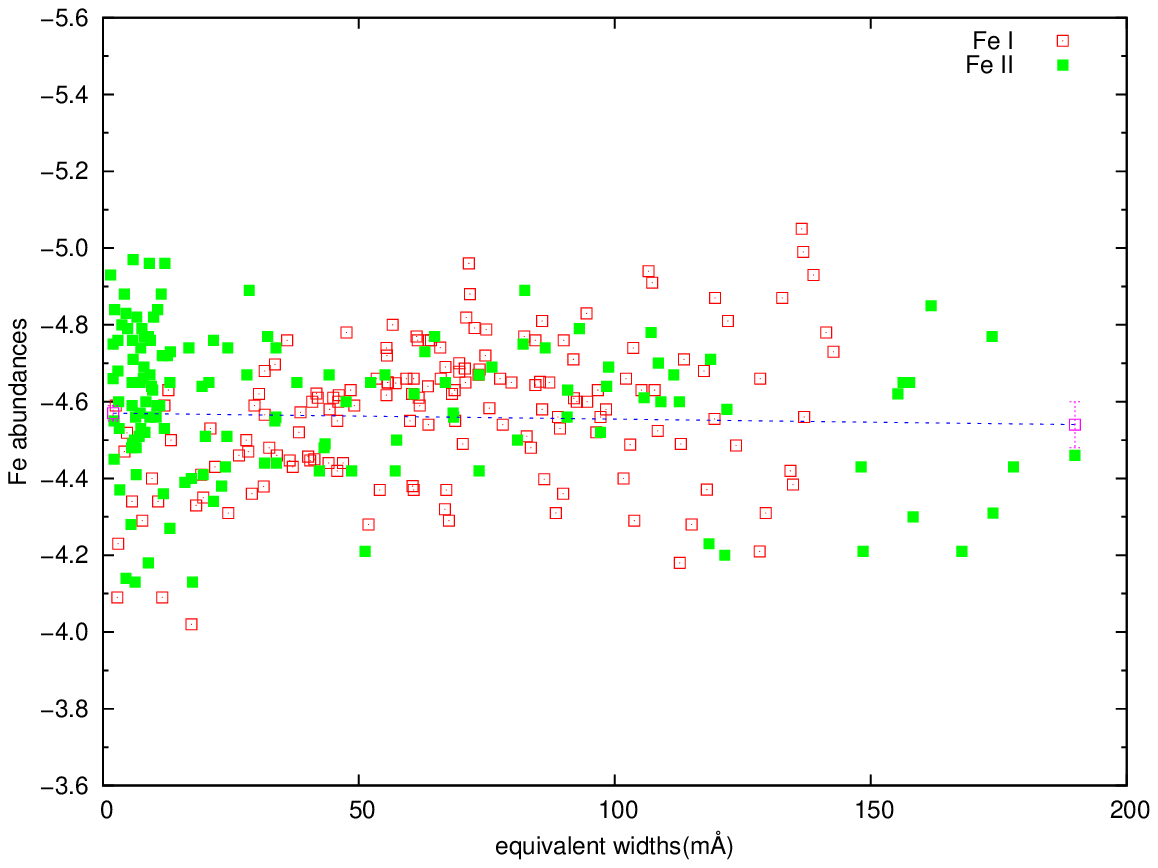}
\caption[]{Fe abundances versus their equivalent widths  and the different excitation potential values are plotted. The correlation between excitation potentials and Fe abundances is minimum at \teff\,=7700 K. \label{fig:EP} } 
\end{figure*}

The SED in Fig.~\ref{fig:sed} was reproduced using \texttt{ATLAS9} flux models. The spectrophotometric data were obtained from \citet{ruban2006}. The photometric data, angular diameter and E(B-V) are given in Table~\ref{table:parameters}. The zero-points reported by \citet{heber2002} were used to transform the various magnitudes into monochromatic fluxes. It was assumed that \textit{y = V} to transform \textit{b-y, c$_{1}$} and \textit{m$_{1}$} indexes to \textit{u, v, b} and \textit{y} magnitudes. The computed fluxes of \astrobj{HD~80404} (\teff = 7700 K and \logg = 1.60) were also consistent with its SED. The synthetic spectrum was reproduced for H$_\delta$ increasing and decreasing  $\pm$ 150\,K or $\pm$ 0.15 dex, \teff, and \logg, respectively using \texttt{SYNTHE}\citep{1981SAOSR.391.....K}. We also used the bluest part of the Balmer series. Different to the synthetic Balmer series spectrum, we also tried to determine the \teff, \logg\ pair of the Paschen series, for which we obtained a good fit. Paschen lines were previously used by \citet{2008A&A...479..849S} to determine the atmospheric parameters of Deneb. The excitation potentials of \ion{Fe}{i} and \ion{Fe}{ii} lines in our abundance analysis ranged from 0.00\,eV to 10\,eV. Their abundances and excitation potentials showed no correlation at \teff = 7700 K (see Fig.~\ref{fig:EP}). The slope of the excitation potential to Fe abundances was -4.943$\times$10$^{-3}$\,$\pm$3.764$\times$10 $^{-3}$ dex $^{-1}$. This is another method to determine the atmospheric parameters, such as \teff\,. Moreover, the ionization equilibrium is also good tool to determine the stellar parameters. Ionization equilibrium for the consecutive ionization stages of \ion{Cr}{i/ii} and \ion{Fe}{i/ii} were fullfilled in the atmospheres of \astrobj{HD\,80404}. The errors in the determined values of \teff\, and \logg\, were assumed to be $\pm$ 150 K and $\pm$ 0.15 dex, as determined from their H${\delta}$ fits, and the error of the microturbulent velocity value was assumed to be 0.7 km s$^{-1}$, as obtained from microturbulence velocity determinations of individual elements in Table ~\ref{table:microturbulence} . 

\section{The results of the abundance analysis}
\begin{figure*}

\center
\includegraphics[width=120mm,height=70mm]{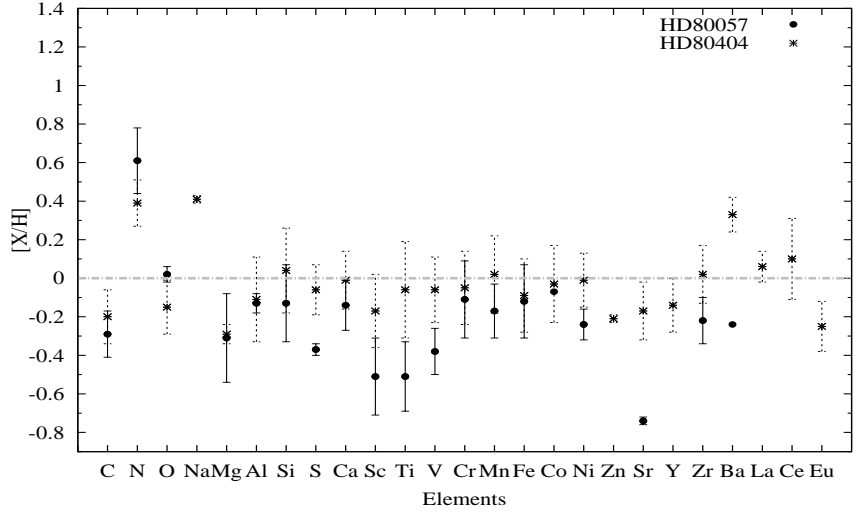}
\caption{The chemical abundances of \astrobj{HD\,80057} and \astrobj{HD\,80404} compared to the solar values in \citet{grevesse1996}.\label{fig:abun}}
\end{figure*}

The elemental abundances found in this study for \astrobj{HD\,80057} and \astrobj{HD\,80404} are presented in Table~\ref{table:abundances} and Fig.~\ref{fig:abun}, together with a comparison of our results with solar composition and previous studies. The systematic error calculated for the abundances of \astrobj{HD\,80404}  are given in Table ~\ref{table:abserr}. The detailed abundances are given in Table \ref{table:appendix},  which also includes the elements used in the analysis, the wavelengths of the identified lines, gf values, and their references. The ionization equilibrium of different elements/ions for target stars are seen in Table~\ref{table:abundances}.

While the errors in \teff\, have the strongest effect on the abundances of \ion{Mg}{ii}, \ion{Al}{i}, \ion{Fe}{i} and \ion{Ba}{ii}, the error in \logg\, affects most strongly those of \ion{Mg}{ii}, \ion{S}{ii}, \ion{Ca}{i} and \ion{Ca}{ii}. This might be due to the fact that these species have the strongest lines with the large EWs.

The sum of CNO abundances of both stars have a solar value. For \astrobj{HD\,80057}, $\alpha$-process elements, except for Si and Ca, are also under-abundant. Al, Sc, Ti, and V are found to be under-abundant. However, Cr, Mn, Fe, and Ni abundances are found to be closer to solar values. Sc, Ti, and V are susceptible to non-LTE effects, which can be ascribed to the value of their second ionization potentials. These are the lowest second ionisation potential in the Iron Group \citep{2002PhDT.........3P}. The heavy elements (Sr, Y, Zr, Ba) are also underabundant with respect to solar abundances.

In the atmospheres of  \astrobj{HD\,80404}; $\alpha$-process elements (Si, S and Ca) and the light element Al are closer to solar values, Mg is deficient, however Na is overabundant.  Sc, Ti, V, Cr, Mn, Fe, Co and Ni  are all closer to solar values in the atmosphere of \astrobj{HD\,80404}. The heavy elements tend to have values slightly smaller than solar, Ba is overabundant\,(see Fig.~\ref{fig:sed}).

\section{Results and Discussion}
\label{Results}

As a result, the [M$/$H] ratio of \astrobj{HD\,80057} is found to be -0.15 $\pm$,\ 0.24 dex when we exclude the over-abundant Na due to NLTE effects \citep{1994PASJ...46..395T, 1995ApJS...99..659V,takeda2008}, the under-abundant Sc and Ti, which are all susceptible to non-LTE effects. The [M$/$H] ratio of \astrobj{HD\,80404} is estimated to be -0.02 $\pm$ 0.20 dex when we exclude the over-abundant Na element. The $N/C$, $N/O$ and $\Sigma$\,$CNO$ values are given in Table~\ref{table:cno}.

\subsection{Evolutionary Status}
\label{evolution}
Light-element, the sum of\,$CNO$ composition reflects mixing process present in the interior of the star. Both stars show a deficiency of C and O, and an enrichment of N. Hence, the combined\,$CNO$ abundances were found to be close to the solar value (see Table~\ref{table:cno}). Both stars also have solar metallicity. The values of $N/C$ and $N/O$ predicted from a linear interpolation of the closest isochrones are consistent with their calculated values (see Table~\ref{table:cno}). The $N/O$ ratio of \astrobj{HD\,80057} is found to be slightly smaller than the theoretical value ~0.52. 

The early CNO contamination in the surface layers of stars can be explained by rotating models. Rotation is an essential factor of stellar models that have a profound effect on the evolution of especially massive stars. Red giants or supergiants, whose progenitor is fast rotating progenitors, rotate six times faster and show N/C ratios three time higher than those formed by  slow rotators \citep{2010A&A...517A..38P, 2013A&A...553A..24G, 2014A&A...565A..39M}. \citet{2013A&A...553A..24G} provide extended data of stellar models including the mass range from 1.7 to 15 M$_{\odot}$  with three different metallicities and with nine different initial rotation velocity models. In the framework of their study, one can see that C, N, and O abundances varies with different initial rotating models. So, we take into account, initial rotational velocities, $N/C$ and $N/O$ ratios, besides \teff\, and \logg\, in stellar evolution models.

In order to investigate the evolutionary states of our targets, we used the Geneva Stellar Model \citep{2013A&A...553A..24G} interactive tools\footnote{http://obswww.unige.ch/Recherche/evoldb/index/} to interpolate between the existing evolutionary tracks that would lead to models with parameters matching those of our targets. We used the relevant parameters (log~T$_{eff}$ and \logg) from \citet{FP2012} for \astrobj{HD~80057} and our own measurements for \astrobj{HD~80404}. 

We experimented by interpolating models for different masses of grids that covered a wide range of masses between 1.7 and 15 M$_{\odot}$.  We kept the metallicity at solar composition (Z = 0.014), and the rotation rate at ${\varv}$ / ${\varv}$\,$_{crit}$ = 0.0 to 0.95 in each case. 

\begin{figure*}
\center
\includegraphics[width=120mm]{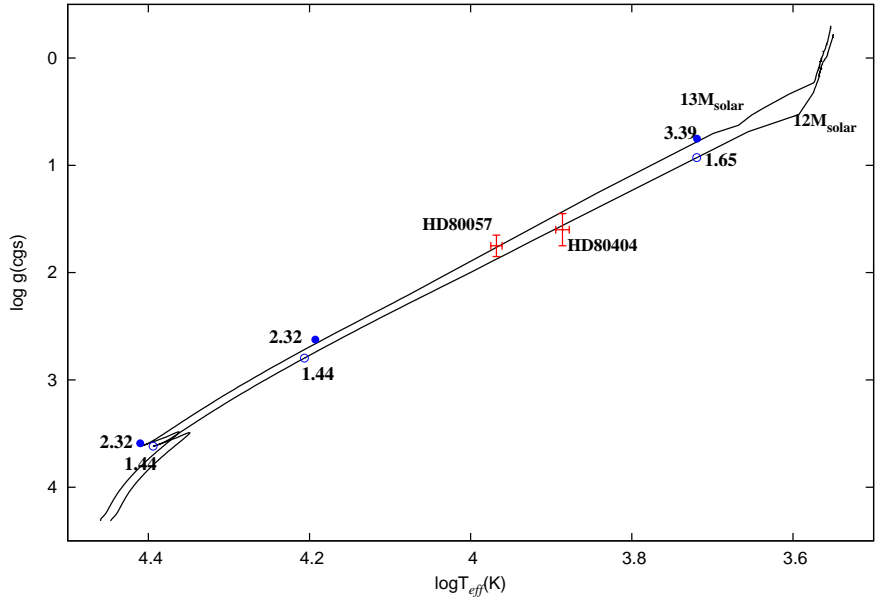}
\caption{The positions of \astrobj{HD\,80057} and \astrobj{HD\,80404} on log~\teff\, \logg\, plane, and evolutionary tracks for 12 and 13 M$_{\odot}$, found by interpolation between existing tracks of Geneva stellar models \citet{2013A&A...553A..24G}. We labelled certain $N/C$ values (by mass ratio) on each of the tracks for comparison with our measurements\label{fig:tracks}.}

\end{figure*}

We searched for the probable evolutionary track, which would be the one with the closest agreement with the atmospheric parameters of our stars (e.g. effective temperatures and surface gravities). We found that both of our stars have masses between 10 M$_{\odot}$ and 14 M$_{\odot}$. While \astrobj{HD~80057} has a mass consistent with the track for 13 M$_{\odot}$, \astrobj{HD~80404} is very close to the track for 12 M$_{\odot}$.  The latter result is based on its fundamental parameters (Fig.~\ref{fig:tracks}), which assumes solar composition and rotation with values of $\Omega$ / $\Omega_{crit}$\,= 0.60, 0.50 (see Table~\ref{table:cno}). These gave globally good fit to stars from the main-sequence to the positions of the blue supergiants before the red supergiant phase in the Hertzsprung-Russell Diagram (hereafter HRD) \citep{2013A&A...553A..24G}.

\begin{figure*}
\center
\includegraphics[width=120mm]{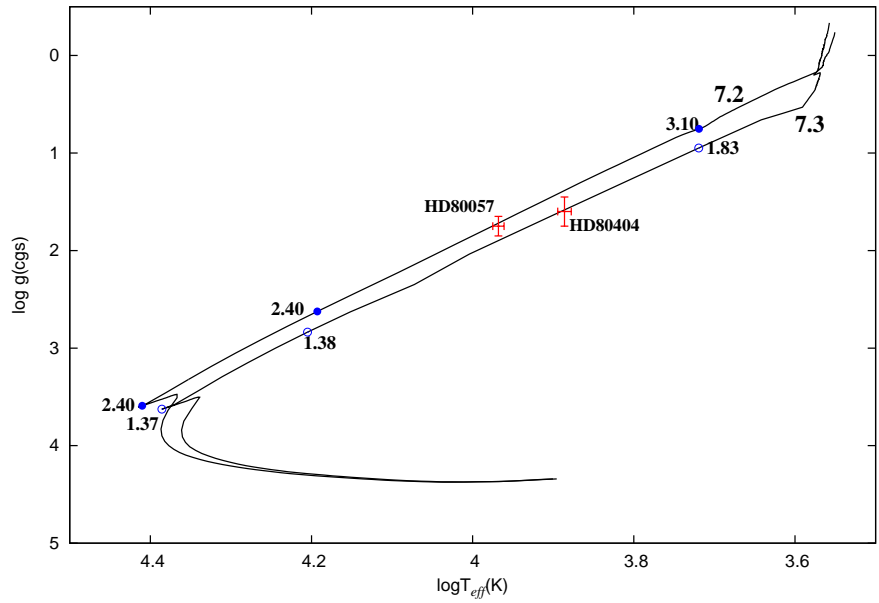}
\caption{The positions of \astrobj{HD\,80057} and \astrobj{HD\,80404} on log~\teff\, - \logg\, plane, and two isochrones (10$^{7.2}$ and 10$^{7.3}$ years) computed by making use of Geneva stellar models \citet{2013A&A...553A..24G} in solid curves. We labelled certain $N/C$ values (by mass ratio) on each of the isochrones for comparison with our measurements.\label{fig:isochrone}}
\end{figure*}

On the log~\teff\, - \logg\, plane, we also computed isochrones using the same online tools and selected the ones to which our stars' parameters had the closest matches (Fig.~\ref{fig:isochrone}). Our results indicate ages close to 16 Myrs for \astrobj{HD~80057} and  19 Myrs for \astrobj{HD~80404}. When the errors in the measurements of the parameters, expressed as the error bars in Fig.~\ref{fig:isochrone}, were considered, a good estimate was obtained for ages between 12.5  and 25 Myrs for both of our stars. The abundance ratios $N/C$ that we computed in this study (2.45 for \astrobj{HD~80057} and 1.57 for \astrobj{HD~80404}) are somewhat consistent (within the uncertainty limits) with the stars' positions on the log\teff\, -- \logg\, plane, for which the expected $N/C$ ratios are printed next to each of the isochrones in Fig.\ref{fig:isochrone}. The derived values of the $N/C$ and $N/O$ ratios from the isochrones are given in Table~\ref{table:cno}. These ratios reveal that $CNO$ mixing processes are active, and that these stars are at the Blue Supergiant (BSG) phase in their evolution prior to the Red Supergiant (RSG) phase, as according to \citet{2013MNRAS.433.1246S}.



\section{Acknowledgements}
\label{acknowledgements}

This research utilised the SIMBAD database, which is operated at CDS, Strasbourg, France.  This work, made use of the MAST-IUE archive (http://archive.stsci.edu/iue/) of SAO/NASA ADS, is based on data obtained from UVES Paranal Observatory Project (ESO DDT Program ID 266.D-5655),  of the VALD database, operated at Uppsala University, the Institute of Astronomy RAS in Moscow, and the University of Vienna. Atomic data compiled in the DREAM data base (E. Biemont, P. Palmeri \& P. Quinet, Astrophys. Space Sci. 269-270, 635, 1999) were extracted via VALD (Kupka et al., 1999, A\&AS 138, 119, and references therein). The authors thank the anonymous referee, whose useful comments helped to improve this work.


\clearpage
\label{figures}

\clearpage


\clearpage


\clearpage
\label{tables}

\begin{table}[ht]

\caption{Stellar parameters of HD\,80057 and HD\,80404 from other authors.}
\label{table:parameters}
\small
\begin{center}
\begin{tabular}{lcc}
\hline
      & HD\,80057 & HD\,80404 \\
\hline
\\
\textbf{Basic}	& &\\
\\
Name	  					& \ldots & \astrobj{$\iota$ Car } \\
Association 				& Vela OB1$^{~a}$&\ldots \\
Spectral type 				& A1-Iab$^{~b}$&A8-Ib$^{~f}$\\
Distance (\,kpc)			& 1.449\,$\pm$\,0.819, 1.839$^{~c,b}$&0.235\,$\pm$\,0.005$^{~c}$\\
Radial velocity (\,km\,s$^{-1}$)		& 25.7\,$\pm$\,2.0$^{~d}$&12.0\,$\pm$\,0.3$^{~d}$ \\[2mm]
\textbf{Atmospheric}	& &\\
\\
\teff\,(K)	 &9300\,$\pm$\,150$^{~b}$&7500\,$\pm$\,200$^{~g}$\\
\logg (cgs)		&1.75\,$\pm$\,0.10$^{~b}$ &2.40\,$\pm$\,0.25$^{~g}$\\
$\xi$ (\,km\,s$^{-1}$)	& 5\,$\pm$\,1$^{~b}$&2.35\,$\pm$\,0.35$^{~g}$\\
$\zeta$ (\,km\,s$^{-1}$)	& 27\,$\pm$\,5$^{~b}$&\ldots\\
$v \sin{i}$ (\,km\,s$^{-1}$)	& 13\,$\pm$\,5$^{~b}$&10.0\,$\pm$\,0.6$^{~d}$\\[2mm]

\\

\textbf{Photometric}			&  & \\
\\
\textbf{(Johnson)}			&  & \\
$V$ 			& 6.$^{m}$044 $\pm$ 0$^{m}$.016$^{~b}$ & 2.$^{m}$26$^{~h}$ \\
$B-V$ 			& 0.$^{m}$285 $\pm$ 0$^{m}$.006$^{~b}$ & 0.$^{m}$18$^{~h}$ \\
$U-B$ 			& -0.$^{m}$117 $\pm$ 0$^{m}$.021$^{~b}$ & 0.$^{m}$16$^{~h}$ \\
\\
\textbf{(Str\"{o}mgren)}			& &\\
$b-y$ 			& 0.267\,$\pm$\,0.003$^{~e}$ &0.123\,$\pm$\,0.007$^{~i}$\\
$m{_1}$ 			& -0.009\,$\pm$\,0.011$^{~e}$ &0.130\,$\pm$\,0.009$^{~i}$\\
$c{_1}$ 			& 0.861\,$\pm$\,0.007$^{~e}$ &1.542\,$\pm$\,0.011$^{~i}$\\
\\
$M_{V}$		& -6.$^{m}$40 $\pm$ 0.$^{m}$18$^{~a}$ & -5.$^{m}$1$^{~j}$\\
$(m - M)_{0}$		& 9.$^{m}$52 $\pm$ 0.$^{m}$18\,$^{~a}$ & 7.$^{m}$2$^{~j}$\\
$M_{\rm bol}$ 		& -6.$^{m}$49 $\pm$ 0.$^{m}$18 \, $^{~a}$ & -5.$^{m}$3$^{~k}$\\
$BC$			& -0.$^{m}$09 & 0.$^{m}$21$^{~k}$\\
$E(B - V)$  		& 0.$^{m}$32 $\pm$ 0.02$^{~b}$ & 0.$^{m}$13$^{~j}$ \\
$\theta{_D}(mas)$  		& \dots &  1.$^{m}$55 $\pm$ 0.12$^{~l}$  \\
\\[2mm]

\hline
\end{tabular}
\end{center}
$~^a$\citet{reed2000},
$~^b$\citet{FP2012},
$~^c$\citet{van2007},
$~^d$\citet{gontcharov2006},
$~^e$\citet{HAU1998},
$~^f$\citet{tetzlaff2011},
$~^g$\citet{smiljanic2006},
$~^h$\citet{ducati2002},
$~^i$\citet{ferromantegazza1996},
$~^j$\citet{snow1994}
$~^k$\citet{1988A&A...195..172V} 
$~^l$\citet{2009MNRAS.394.1620D}

\end{table}
\begin{sidewaystable*}[ht!]
\small
\caption{Atmospheric parameters of \astrobj{HD\,80057} and 
\label{table:literature}
\astrobj{HD\,80404} from various sources.}
\setlength{\tabcolsep}{.05cm}
\centering
\begin{tabular}{l c c c c c c}
\hline\hline
Source						& \teff (K)	& \logg	& $\xi$(km\,s$^{-1}$)	& $\zeta$	(km\,s$^{-1}$) &Spectra &Method\\
\hline
\\
\multicolumn{5}{c}{{\underline{     HD 80057     }}}\\[3mm] 
\citet{FP2012}   &9300\,$\pm$\,150	& 1.75\,$\pm$0.10	& 5\,$\pm$\,1 &27 &FEROS, R$\sim$40000 &NLTE \\[2mm]
   && 	&  & &3500-9300 \AA,\, S/N $>$ 150 & \\[2mm]
\\
\multicolumn{5}{c}{{\underline{     iota Car    }}}\\[3mm]
\citet{boiarchuk1984}&7300\,$\pm$\,200  	            & 1.40$\pm$0.10	            & \ldots         &\ldots  &reciprocal dispersion, 2 \AA\ /mm &LTE, \citet{kurucz1979}\\[2mm]
\citet{luck1985}&7500  	            & 0.90	            & 2.5         &1  &&LTE, Fe I/\,II  \\[2mm]
 &   	            &  	            &           &   && ionization balance \\[2mm]
\citet{luck1992}&7500  	            & 1.6	            & 2.2         &1  & CTIO, R$\sim$18000&LTE\\[2mm]
 
&&&&&\\[2mm]
\citet{smiljanic2006}&7500  	            & 2.40	            & 2.34         &\ldots  &FEROS, R=48000& LTE, $H_\alpha$, and Fe I/\,II \\[2mm]
 &   	            &  	            &           &   &3500-9200 \AA,\, S/N $>$ 150& ionization balance \\[2mm]
\hline
\end{tabular}
\end{sidewaystable*}
\begin{table}[ht]
\setlength{\tabcolsep}{.10cm}
\small
\centering
\caption{The comparison of derived abundances of target stars 
\label{table:abundances}
 relative to the solar values and literature.} 
\begin{tabular}{cllllllllll}
\hline
\multicolumn{1}{c}{} &Solar$^1$ &\multicolumn{4}{c}{\astrobj{HD\,80057}} &\multicolumn{5}{c}{\astrobj{HD\,80404}}\\
\multicolumn{1}{c}{Species}&\multicolumn{1}{c}{} &\multicolumn{1}{c}{n}   &\multicolumn{1}{c}{This Study}    &\multicolumn{2}{c}{FP$^{2,3}$}    &\multicolumn{1}{c}{n}   &\multicolumn{1}{c}{This Study} &\multicolumn{1}{c}{n}  & \multicolumn{2}{c}{LL$^{4,5}$}      \\ 
\multicolumn{1}{c}{}&\multicolumn{1}{c}{} &\multicolumn{2}{c}{\hrulefill}       &\multicolumn{2}{c}{}    &\multicolumn{2}{c}{\hrulefill}    &\multicolumn{1}{c}{}  & \multicolumn{2}{c}{}      \\               


\hline

\ion{C}{i}	&	-3.45	&	\multicolumn{1}{c}{5}    	&	\multicolumn{1}{c}{-3.72$\pm$0.22} 	&	\multicolumn{2}{c}{-3.78$\pm$0.08} 	&	\multicolumn{1}{c}{12} 	&	\multicolumn{1}{c}{-3.65$\pm$0.14}  	&	\multicolumn{1}{c}{\ldots}   	&	\multicolumn{2}{c}{-3.67}     \\
\ion{C}{ii}	&	-3.45	&	\multicolumn{1}{c}{2}    	&	\multicolumn{1}{c}{-3.78$\pm$0.04} 	&	\multicolumn{2}{c}{-3.72$\pm$0.13} 	&	\multicolumn{1}{c}{\ldots} 	&	\multicolumn{1}{c}{\ldots}  	&	\multicolumn{1}{c}{\ldots}   	&	\multicolumn{2}{c}{\ldots}     \\ 
\ion{N}{i}	&	-4.03	&	\multicolumn{1}{c}{1}    	&	\multicolumn{1}{c}{-3.42$\pm$0.17} 	&	\multicolumn{2}{c}{-3.66$\pm$0.04} 	&	\multicolumn{1}{c}{3} 	&	\multicolumn{1}{c}{-3.52$\pm$0.14}  	&	\multicolumn{1}{c}{\ldots}   	&	\multicolumn{2}{c}{-3.72}     \\  
\ion{N}{ii}	&	-4.03	&	\multicolumn{1}{c}{\ldots}    	&	\multicolumn{1}{c}{\ldots} 	&	\multicolumn{2}{c}{-3.71} 	&	\multicolumn{1}{c}{\ldots} 	&	\multicolumn{1}{c}{\ldots}  	&	\multicolumn{1}{c}{\ldots}   	&	\multicolumn{2}{c}{\ldots}     \\   
\ion{O}{i} 	&	-3.13	&	\multicolumn{1}{c}{3}    	&	\multicolumn{1}{c}{-3.11$\pm$0.04} 	&	\multicolumn{2}{c}{\ldots} 	&	\multicolumn{1}{c}{14} 	&	\multicolumn{1}{c}{-3.34$\pm$0.24}  	&	\multicolumn{1}{c}{3}   	&	\multicolumn{2}{c}{-3.38}     \\  

\ion{Na}{i}	&	-5.67     	&	\multicolumn{1}{c}{\ldots}    	&	\multicolumn{1}{c}{\ldots} 	&	\multicolumn{2}{c}{\ldots} 	&	\multicolumn{1}{c}{3} 	&	\multicolumn{1}{c}{-5.26$\pm$0.02}  	&	\multicolumn{1}{c}{2}   	&	\multicolumn{2}{c}{-5.19$\pm$0.03}     \\
\ion{Mg}{i}	&	-4.42	&	\multicolumn{1}{c}{2}    	&	\multicolumn{1}{c}{-4.78$\pm$0.22} 	&	\multicolumn{2}{c}{-4.58$\pm$0.06} 	&	\multicolumn{1}{c}{1} 	&	\multicolumn{1}{c}{-4.67}  	&	\multicolumn{1}{c}{10}   	&	\multicolumn{2}{c}{-4.35$\pm$0.23}     \\   

\ion{Mg}{ii}	&	-4.42	&	\multicolumn{1}{c}{5}    	&	\multicolumn{1}{c}{-4.71$\pm$0.08} 	&	\multicolumn{2}{c}{-4.60$\pm$0.06} 	&	\multicolumn{1}{c}{4} 	&	\multicolumn{1}{c}{-4.72$\pm$0.05}  	&	\multicolumn{1}{c}{\ldots}   	&	\multicolumn{2}{c}{\ldots}     \\  
\ion{Al}{i}&	-5.53	&	\multicolumn{1}{c}{2}    	&	\multicolumn{1}{c}{-6.08$\pm$0.05} 	&	\multicolumn{2}{c}{\ldots} 	&	\multicolumn{1}{c}{2} 	&	\multicolumn{1}{c}{-5.56$\pm$0.22}  	&	\multicolumn{1}{c}{1}   	&	\multicolumn{2}{c}{-5.58}     \\ 
\ion{Al}{ii}	&	-5.53	&	\multicolumn{1}{c}{1}    	&	\multicolumn{1}{c}{-6.03} 	&	\multicolumn{2}{c}{\ldots} 	&	\multicolumn{1}{c}{1} 	&	\multicolumn{1}{c}{-5.81}  	&	\multicolumn{1}{c}{\ldots}   	&	\multicolumn{2}{c}{\ldots}     \\ 
\ion{Si}{i}	&	-4.45	&	\multicolumn{1}{c}{\ldots}    	&	\multicolumn{1}{c}{\ldots} 	&	\multicolumn{2}{c}{\ldots} 	&	\multicolumn{1}{c}{2} 	&	\multicolumn{1}{c}{-4.16$\pm$0.00}  	&	\multicolumn{1}{c}{27}   	&	\multicolumn{2}{c}{-4.23$\pm$0.16}     \\  
\ion{Si}{ii}	&	-4.45	&	\multicolumn{1}{c}{11}    	&	\multicolumn{1}{c}{-4.57$\pm$0.20} 	&	\multicolumn{2}{c}{\ldots} 	&	\multicolumn{1}{c}{7} 	&	\multicolumn{1}{c}{-4.66$\pm$0.22}  	&	\multicolumn{1}{c}{\ldots}   	&	\multicolumn{2}{c}{\ldots}     \\ 
\ion{S}{i} 	&	-4.67	&	\multicolumn{1}{c}{\ldots}    	&	\multicolumn{1}{c}{\ldots} 	&	\multicolumn{2}{c}{\ldots} 	&	\multicolumn{1}{c}{3} 	&	\multicolumn{1}{c}{-4.63$\pm$0.08}  	&	\multicolumn{1}{c}{9}   	&	\multicolumn{2}{c}{-4.63$\pm$0.14}     \\   
\ion{S}{ii}	&	-4.67	&	\multicolumn{1}{c}{2}    	&	\multicolumn{1}{c}{-5.04$\pm$0.03}	&	\multicolumn{2}{c}{\ldots} 	&	\multicolumn{1}{c}{3} 	&	\multicolumn{1}{c}{-4.82$\pm$0.10}  	&	\multicolumn{1}{c}{\ldots}   	&	\multicolumn{2}{c}{\ldots}     \\   
\ion{Ca}{i}	&	-5.64	&	\multicolumn{1}{c}{1}    	&	\multicolumn{1}{c}{-5.75} 	&	\multicolumn{2}{c}{\ldots} 	&	\multicolumn{1}{c}{24} 	&	\multicolumn{1}{c}{-5.64$\pm$0.15}  	&	\multicolumn{1}{c}{12}   	&	\multicolumn{2}{c}{-5.57$\pm$0.17}     \\    
\ion{Ca}{ii}	&	-5.64	&	\multicolumn{1}{c}{2}    	&	\multicolumn{1}{c}{-5.79$\pm$0.16} 	&	\multicolumn{2}{c}{\ldots} 	&	\multicolumn{1}{c}{3} 	&	\multicolumn{1}{c}{-5.75$\pm$0.02}  	&	\multicolumn{1}{c}{3}   	&	\multicolumn{2}{c}{-5.54$\pm$0.40}     \\ 
\ion{Sc}{ii}	&	-8.83	&	\multicolumn{1}{c}{7}    	&	\multicolumn{1}{c}{-9.34$\pm$0.20} 	&	\multicolumn{2}{c}{\ldots} 	&	\multicolumn{1}{c}{26} 	&	\multicolumn{1}{c}{-9.00$\pm$0.20}  	&	\multicolumn{1}{c}{5}   	&	\multicolumn{2}{c}{-9.05$\pm$0.13}     \\ 
\ion{Ti}{i}	&	-6.98	&	\multicolumn{1}{c}{\ldots}    	&	\multicolumn{1}{c}{\ldots} 	&	\multicolumn{2}{c}{\ldots} 	&	\multicolumn{1}{c}{40} 	&	\multicolumn{1}{c}{-6.89$\pm$0.24}  	&	\multicolumn{1}{c}{\ldots}   	&	\multicolumn{2}{c}{\ldots}     \\ 
\ion{Ti}{ii}	&	-6.98	&	\multicolumn{1}{c}{43}    	&	\multicolumn{1}{c}{-7.49$\pm$0.18} 	&	\multicolumn{2}{c}{\ldots} 	&	\multicolumn{1}{c}{59} 	&	\multicolumn{1}{c}{-7.15$\pm$0.21}  	&	\multicolumn{1}{c}{1}   	&	\multicolumn{2}{c}{-7.05}     \\ 
\ion{V}{i}	&	-8.00	&	\multicolumn{1}{c}{\ldots}    	&	\multicolumn{1}{c}{\ldots} 	&	\multicolumn{2}{c}{\ldots} 	&	\multicolumn{1}{c}{4} 	&	\multicolumn{1}{c}{-8.05$\pm$0.11}  	&	\multicolumn{1}{c}{\ldots}   	&	\multicolumn{2}{c}{\ldots}     \\ 
\ion{V}{ii}	&	-8.00	&	\multicolumn{1}{c}{7}    	&	\multicolumn{1}{c}{-8.38$\pm$0.12} 	&	\multicolumn{2}{c}{\ldots} 	&	\multicolumn{1}{c}{25} 	&	\multicolumn{1}{c}{-8.06$\pm$0.18}  	&	\multicolumn{1}{c}{\ldots}   	&	\multicolumn{2}{c}{\ldots}     \\ 
\ion{Cr}{i}	&	-6.33	&	\multicolumn{1}{c}{\ldots}    	&	\multicolumn{1}{c}{\ldots} 	&	\multicolumn{2}{c}{\ldots} 	&	\multicolumn{1}{c}{40} 	&	\multicolumn{1}{c}{-6.40$\pm$0.19}  	&	\multicolumn{1}{c}{\ldots}   	&	\multicolumn{2}{c}{\ldots}     \\ 
\ion{Cr}{ii}	&	-6.33	&	\multicolumn{1}{c}{50}    	&	\multicolumn{1}{c}{-6.44$\pm$0.17} 	&	\multicolumn{2}{c}{\ldots} 	&	\multicolumn{1}{c}{59} 	&	\multicolumn{1}{c}{-6.37$\pm$0.18}  	&	\multicolumn{1}{c}{\ldots}   	&	\multicolumn{2}{c}{\ldots}     \\    
\ion{Mn}{i}&	-6.61	&	\multicolumn{1}{c}{\ldots}    	&	\multicolumn{1}{c}{\ldots} 	&	\multicolumn{2}{c}{\ldots} 	&	\multicolumn{1}{c}{25} 	&	\multicolumn{1}{c}{-6.56$\pm$0.22}  	&	\multicolumn{1}{c}{\ldots}   	&	\multicolumn{2}{c}{\ldots}     \\
\ion{Mn}{ii}	&	-6.61	&	\multicolumn{1}{c}{8}    	&	\multicolumn{1}{c}{-6.78$\pm$0.14} 	&	\multicolumn{2}{c}{\ldots} 	&	\multicolumn{1}{c}{11} 	&	\multicolumn{1}{c}{-6.66$\pm$0.11}  	&	\multicolumn{1}{c}{\ldots}   	&	\multicolumn{2}{c}{\ldots}     \\   
\ion{Fe}{i}	&	-4.50	&	\multicolumn{1}{c}{25}    	&	\multicolumn{1}{c}{-4.70$\pm$0.21} 	&	\multicolumn{2}{c}{\ldots} 	&	\multicolumn{1}{c}{161} 	&	\multicolumn{1}{c}{-4.59$\pm$0.18}  	&	\multicolumn{1}{c}{61}   	&	\multicolumn{2}{c}{-4.29$\pm$0.19}     \\  
\ion{Fe}{ii}	&	-4.50	&	\multicolumn{1}{c}{105}    	&	\multicolumn{1}{c}{-4.62$\pm$0.18} 	&	\multicolumn{2}{c}{\ldots} 	&	\multicolumn{1}{c}{139} 	&	\multicolumn{1}{c}{-4.58$\pm$0.20}  	&	\multicolumn{1}{c}{12}   	&	\multicolumn{2}{c}{-4.30$\pm$0.16}     \\ 
\ion{Fe}{iii}	&	-4.50	&	\multicolumn{1}{c}{3}    	&	\multicolumn{1}{c}{-4.69$\pm$0.06} 	&	\multicolumn{2}{c}{\ldots} 	&	\multicolumn{1}{c}{\ldots} 	&	\multicolumn{1}{c}{\ldots}  	&	\multicolumn{1}{c}{\ldots}   	&	\multicolumn{2}{c}{\ldots}     \\ 
\ion{Co}{i} 	&	-7.08	&	\multicolumn{1}{c}{\ldots}    	&	\multicolumn{1}{c}{\ldots} 	&	\multicolumn{2}{c}{\ldots} 	&	\multicolumn{1}{c}{7} 	&	\multicolumn{1}{c}{-7.17$\pm$0.15}  	&	\multicolumn{1}{c}{\ldots}   	&	\multicolumn{2}{c}{\ldots}     \\ 

\ion{Co}{ii} 	&	-7.08	&	\multicolumn{1}{c}{1}    	&	\multicolumn{1}{c}{-7.15} 	&	\multicolumn{2}{c}{\ldots} 	&	\multicolumn{1}{c}{5} 	&	\multicolumn{1}{c}{-6.95$\pm$0.20}  	&	\multicolumn{1}{c}{\ldots}   	&	\multicolumn{2}{c}{\ldots}     \\

\ion{Ni}{i}&	-5.75	&	\multicolumn{1}{c}{\ldots}    	&	\multicolumn{1}{c}{\ldots} 	&	\multicolumn{2}{c}{\ldots} 	&	\multicolumn{1}{c}{17} 	&	\multicolumn{1}{c}{-5.73$\pm$0.12}  	&	\multicolumn{1}{c}{\ldots}   	&	\multicolumn{2}{c}{\ldots}     \\ 
\ion{Ni}{ii}	&	-5.75	&	\multicolumn{1}{c}{3}    	&	\multicolumn{1}{c}{-5.99$\pm$0.08} 	&	\multicolumn{2}{c}{\ldots} 	&	\multicolumn{1}{c}{3} 	&	\multicolumn{1}{c}{-5.92$\pm$0.08}  	&	\multicolumn{1}{c}{\ldots}   	&	\multicolumn{2}{c}{\ldots}     \\ 
\ion{Zn}{i}	&	-7.40	&	\multicolumn{1}{c}{\ldots}    	&	\multicolumn{1}{c}{\ldots} 	&	\multicolumn{2}{c}{\ldots} 	&	\multicolumn{1}{c}{2} 	&	\multicolumn{1}{c}{-7.61$\pm$0.02}  	&	\multicolumn{1}{c}{\ldots}   	&	\multicolumn{2}{c}{\ldots}     \\
\ion{Sr}{ii}	&	-9.03	&	\multicolumn{1}{c}{2}    	&	\multicolumn{1}{c}{-9.77$\pm$0.02} 	&	\multicolumn{2}{c}{\ldots} 	&	\multicolumn{1}{c}{4} 	&	\multicolumn{1}{c}{-9.20$\pm$0.15}  	&	\multicolumn{1}{c}{\ldots}   	&	\multicolumn{2}{c}{\ldots}     \\ 
\ion{Y}{ii}	&	-9.76	&	\multicolumn{1}{c}{\ldots}    	&	\multicolumn{1}{c}{\ldots} 	&	\multicolumn{2}{c}{\ldots} 	&	\multicolumn{1}{c}{15} 	&	\multicolumn{1}{c}{-9.90$\pm$0.10}  	&	\multicolumn{1}{c}{1}   	&	\multicolumn{2}{c}{-9.61$\pm$0.03}     \\   
\ion{Zr}{ii}	&	-9.40	&	\multicolumn{1}{c}{3}    	&	\multicolumn{1}{c}{-9.62$\pm$0.12} 	&	\multicolumn{2}{c}{\ldots} 	&	\multicolumn{1}{c}{17} 	&	\multicolumn{1}{c}{-9.42$\pm$0.15}  	&	\multicolumn{1}{c}{\ldots}   	&	\multicolumn{2}{c}{\ldots}     \\ 
\ion{Ba}{ii}	&	-9.87	&	\multicolumn{1}{c}{1}    	&	\multicolumn{1}{c}{-10.11} 	&	\multicolumn{2}{c}{\ldots} 	&	\multicolumn{1}{c}{4} 	&	\multicolumn{1}{c}{-9.54$\pm$0.09}  	&	\multicolumn{1}{c}{2}   	&	\multicolumn{2}{c}{-9.40$\pm$0.09}     \\ 
\ion{La}{ii}	& -10.83	&	\multicolumn{1}{c}{\ldots}    	&	\multicolumn{1}{c}{\ldots} 	&	\multicolumn{2}{c}{\ldots} 	&	\multicolumn{1}{c}{12} 	&	\multicolumn{1}{c}{-10.77$\pm$0.08}  	&	\multicolumn{1}{c}{\ldots}   	&	\multicolumn{2}{c}{\ldots}     \\
\ion{Ce}{ii} 	&	-10.42	&	\multicolumn{1}{c}{\ldots}    	&	\multicolumn{1}{c}{\ldots} 	&	\multicolumn{2}{c}{\ldots} 	&	\multicolumn{1}{c}{9} 	&	\multicolumn{1}{c}{-10.32$\pm$0.21}  	&	\multicolumn{1}{c}{\ldots}   	&	\multicolumn{2}{c}{\ldots}     \\ 
\ion{Eu}{ii}	&	-11.49	&	\multicolumn{1}{c}{\ldots}    	&	\multicolumn{1}{c}{\ldots} 	&	\multicolumn{2}{c}{\ldots} 	&	\multicolumn{1}{c}{4} 	&	\multicolumn{1}{c}{-11.74$\pm$0.13}  	&	\multicolumn{1}{c}{2}   	&	\multicolumn{2}{c}{-11.12$\pm$0.02}     \\ 
\hline
\teff (\,K)	&	 	&	\multicolumn{1}{c}{\ldots}    	&	\multicolumn{1}{c}{9300} 	&	\multicolumn{2}{c}{\ldots} 	&	\multicolumn{1}{c}{ } 	&	\multicolumn{1}{c}{7700}  	&	\multicolumn{1}{c}{ }   	&	\multicolumn{2}{c}{ }     \\ 

\logg(cgs)	&	 	&	\multicolumn{1}{c}{\ldots}    	&	\multicolumn{1}{c}{1.75} 	&	\multicolumn{2}{c}{\ldots} 	&	\multicolumn{1}{c}{ } 	&	\multicolumn{1}{c}{1.60}  	&	\multicolumn{1}{c}{ }   	&	\multicolumn{2}{c}{}     \\ 
\hline
\multicolumn{5}{l}{1. \citet{grevesse1996}, 2. \citet{FP2012}, }\\
\multicolumn{5}{l}{3. \citet{1998SSRv...85..161G}, 4. \citet{luck1992}, 5. \citet{grevesse1984}}\\
\end{tabular}
\end{table}  


\begin{table}
\small
\centering
\caption{Microturbulence determinations from various elements/ions}
\label{table:microturbulence}
\begin{tabular}{llllllll}
\hline
 Element & n    & $\xi_1${\small (scatter) }        & log (N/N$_T$)      & $\xi_2$ {\small (slope)}        & log (N/N$_T$)  & Reference     \\  
         &      & {\small km s $^{-1}$}           &                    &  {\small km s$^{-1}$}         &                &               \\
\hline

\multicolumn{7}{c} {\astrobj{HD\,80057}} \\
\multicolumn{7}{c} {} \\
 Fe II   & 105      & 4.3                   & -4.62$\pm$0.18     &  4.4               & -4.62$\pm$0.18  & KX+ N4  \\       
  \multicolumn{2}{c} {avg}  & 4.4                   &                    &                      &                 &      +VALD        \\
       \multicolumn{2}{c} {stdev}   & 0.1             &                    &                      &                &               \\
                                                                              
 \multicolumn{7}{c} {} \\

\multicolumn{7}{c} {\astrobj{HD\,80404}} \\
\multicolumn{7}{c} {} \\ 
 C I    & 13        & 4.1                    & -3.71$\pm$0.15     &  4.0                & -3.71$\pm$0.15 & WF            \\       
       \multicolumn{2}{c} {adopted}   & 4.1                    &                    &                      &                &               \\
        &           &                         &                    &                      &                &               \\  

 Ca I   & 39        & 2.2                    & -5.67$\pm$0.12     &  2.2                & -5.67$\pm$0.12 & FW+WS         \\       
       \multicolumn{2}{c} {adopted}   & 2.2                    &                    &                      &                &               \\
        &           &                         &                    &                      &                &               \\  
 Fe I   & 152       & 2.0                    & -4.56$\pm$0.16     &  2.0                & -4.56$\pm$0.16 & KX+ N4   \\       
       \multicolumn{2}{c} {adopted}   & 2.0                    &                    &                      &                &             +VALD  \\
        &          &                         &                    &                      &                &                \\ 
 Fe II  & 139       & 2.3                    & -4.63$\pm$0.17     &  2.4                & -4.59$\pm$0.18 & KX+N4    \\       
       \multicolumn{2}{c} {adopted}   & 2.4                    &                    &                      &                &            +VALD   \\
        &          &                         &                    &                      &                &                \\  
 Cr I   & 40        & 1.7                    & -6.36$\pm$0.18     &  1.7                & -6.36$\pm$0.18 & MF            \\
       \multicolumn{2}{c} {adopted}   & 1.7                    &                    &                      &                &               \\
             
        &           &                         &                    &                      &                &               \\ 
 Cr II  & 58        & 2.0                    & -6.32$\pm$0.17     &  2.0                & -6.32$\pm$0.17 & MF+KX      \\
       \multicolumn{2}{c} {adopted}   & 2.0                    &                    &                      &                &            +NL   \\
        &           &                         &                    &                      &                &               \\  
 Sr II  & 4         & 1.8                    & -9.02$\pm$0.10     &  1.9                & -9.02$\pm$0.10 & B+WM          \\
       \multicolumn{2}{c} {adopted}   & 1.9                    &                    &                      &                &               \\
        &           &                         &                    &                      &                &               \\  
  Y II  & 15        & 1.7                    & -9.77$\pm$0.13     &  1.8                & -9.77$\pm$0.14 & HL            \\

       \multicolumn{2}{c} {adopted}   & 1.7                    &                    &                      &                &               \\
        &           &                         &                    &                      &                &               \\  
Zr II   & 17      & 1.7                    & -9.33$\pm$0.12     &  1.7                & -9.33$\pm$0.12 & LN+BG         \\
       \multicolumn{2}{c} {adopted}   & 1.7                   &                    &                      &                &               \\
        &           &                         &                    &                      &                &               \\
       \multicolumn{2}{c} {avg}   & 2.2             &                    &                      &                &               \\
       \multicolumn{2}{c} {stdev}   & 0.7             &                    &                      &                &               \\
\hline

\hline
\multicolumn{7}{l}{References of gf-values:B= \citet{1998ApJ...496.1051B}}; BG = \citet{biemont1981}; \\ 
\multicolumn{7}{l}{HL = \citet{hannaford1982};  }\\
\multicolumn{7}{l}{FW = \cite{fuhr2002} and \citet{1988JPCRD..17S....F};  }\\

\multicolumn{7}{l}{KX = \citet{1995ASPC...78..205K}, LN = \citet{ljung2006};}\\

\multicolumn{7}{l}{NL = \citet{nil2006}, N4 = \citet{fuhr2006}, }\\
\multicolumn{7}{l}{MF = \citet{1988JPCRD..17S....F} and \citet{1988atps.book.....M};}\\

\multicolumn{7}{l}{WF = \cite{wiese1996}; WM = \citet{wiese1980}, WS = \citet{wiese1969};}\\
\multicolumn{7}{l}{VALD,VALD2 data=  \citet{1995A&AS..112..525P}, \citet{1997BaltA...6..244R},}\\

\multicolumn{7}{l}{\citet{1999A&AS..138..119K}, \citet{2000BaltA...9..590K};}\\

\end{tabular}\\
\end{table}
\begin{table}
	\centering
	\caption{Error reasons for the abundances of \astrobj{HD\,80404}}
	\setlength{\tabcolsep}{2.0pt}
	\def\arraystretch{0.90}

		\begin{minipage}{85mm}
			\label{table:abserr}
			\begin{tabular}{@{}lccccr@{}}
			\hline
			\hline
Species	& $\sigma$ (\teff)& $\Delta$ (\logg)	& $\Delta$ ($\xi$)		&EW& $\sigma_{TOTAL}$   \\
			& {\footnotesize (+ 150K)}			   &{\footnotesize (+ 0.15 dex)}	&{\footnotesize (+ 0.7 km\,s$^{-1}$)}	& {\footnotesize (10$\%$EW)	}		\\
				\hline

				C I		& 0.06		&-0.04	&-0.02	&0.07	&0.10		\\
				N I		& 0.06		& 0.01	& 0.04	&0.08	&0.08		\\
				O I		& 0.00		& 0.01	&-0.02	&0.04	&0.04		\\
				Na I	& 0.11		&-0.07	&-0.03	&0.06	&0.14		\\
				Mg I	& 0.13		&-0.08	&-0.14	&0.02	&0.20		\\
				Mg II	& 0.00		&0.02	&-0.09	&0.03	&0.10		\\
				Al I	& 0.15		&-0.07	&-0.31	&-0.13	&0.37		\\
				Al II	&-0.05		&0.07	&-0.02	&0.03	&0.09		\\
				Si I	& 0.12		&-0.07	&-0.01	&0.04	&0.15		\\
				Si II	&-0.02		& 0.04	&-0.10	&0.01	&0.11		\\
				S I		& 0.11		&-0.07	&-0.01	&0.05	&0.14		\\
				S II	&-0.10		& 0.18	&-0.02	&0.06	&0.19		\\				  			
				Ca I	& 0.17		&-0.10	&-0.05	&0.07	&0.22		\\
				Ca II	& 0.06		&-0.10	&-0.07	&0.08	&0.12		\\
				Sc II	& 0.10		&-0.01	&-0.14	&0.12	&0.21		\\
				Ti I	& 0.16		&-0.08	&-0.01	&0.00	&0.19		\\
				Ti II	& 0.10		& 0.02	&-0.16	&0.16	&0.25		\\
				V I		& 0.16		&-0.08	&-0.01	&0.04	&0.18		\\
				V II	& 0.08		& 0.00	&-0.09	&0.09	&0.15		\\
				Cr I	& 0.16		&-0.07	&-0.04	&0.04	&0.19		\\
				Cr II	& 0.05		& 0.01	&-0.11	&0.10	&0.16		\\
				Mn I	& 0.14		&-0.08	&-0.05	&0.07	&0.18 		\\
				Mn II	& 0.03		& 0.01	&-0.03	&0.05	&0.07 		\\
				Fe I	& 0.15		&-0.09	&-0.13	&0.13	&0.25		\\
				Fe II	& 0.06		& 0.06	&-0.05	&0.13	&0.16 		\\
				Co I	& 0.17		&-0.07	&-0.03	&0.06	&0.20		\\
				Co II	& 0.02		&-0.04	&-0.17	&0.06	&0.19		\\
				Ni I	& 0.13		&-0.08	&-0.02	&0.05	&0.16		\\
				Ni II	& 0.04		& 0.02	&-0.04	&0.07	&0.09  		\\
				Zn I	& 0.14		&-0.07	& 0.00	&0.05	&0.16  		\\
				Sr II	& 0.17		&-0.04	&-0.33	&0.18	&0.41		\\
				Y II	&0.04		&-0.01	&-0.10	&0.10	&0.15  		\\
				Zr II	&0.03		& 0.00	& 0.06	&0.08	&0.10		\\							
				Ba II	&-0.08		& 0.12	&-0.30	&0.19	&0.38		\\
				La II	&0.14		&-0.04	&-0.03	&0.06	&0.16		\\
				Ce II	&0.13		&-0.03	&-0.02	&0.05	&0.14		\\								
				Eu II	&0.17		&-0.04	&-0.02	&0.05	&0.18		\\
			
				\hline
			
\multicolumn{6}{l}{{\footnotesize $\sigma^2_{tot.}$=$\sigma^2_{teff}$+$\sigma^2_{logg}$+$\sigma^2_{EW}$+$\sigma^2_{\xi}$}}				

			\end{tabular}
			
			\end{minipage}
\end{table}
\begin{table*} 
\scriptsize                                                                     
\caption[ ]{Elemental Abundances of \astrobj{HD\,80057} and \astrobj{HD\,80404}\label{table:appendix}}                      

\begin{flushleft}                                                                   
    \begin{tabular}{cccccrcrc}  
    \hline                                                    
 & & & & &\multicolumn{2}{c}{\astrobj{HD\,80057}}&\multicolumn{2}{c}{\astrobj{HD\,80404}}\\         
Species&Multiplet& $\lambda$(\AA)& log gf&Ref.&W$_{\lambda}$(m{\AA})                
&log N/N$_{T}$ &  W$_{\lambda}$(m{\AA})& log N/N$_{T}$ \\                           
 \hline 
\noalign{\smallskip}

C  I & & & & \multicolumn{3}{c}{log C/N$_{T}$ =-3.72$\pm$0.22} & &-3.65$\pm$0.14\\ 
                                                  
    \noalign{\smallskip}                                                                                                                                                                                                                                                         
                                                                

& 6  & 4771.74 & -1.87 & FW & \ldots  &  \ldots &   74.4  & -3.47\\                       
&    & 4775.90 & -2.90 & FW & \ldots  &  \ldots &   37.7  & -3.47\\                       
& 13 & 5052.17 & -1.30 & CR & \ldots  &  \ldots &   69.9  & -3.90\\                       
&    & 4932.05 & -1.66 & FW & \ldots  &  \ldots &   47.2  & -3.84\\                       
& 14 & 4371.27 & -1.96 & FW & \ldots  &  \ldots &   31.8  & -3.75\\                       
& 17 & 4228.32 & -2.27 & FW & \ldots  &  \ldots &   20.9  & -3.64\\                       
& 26 & 7111.47 & -1.09 & FW & 7.0     &  -3.44  &   42.8  & -3.72\\                                                                                    
&    & 7113.18 & -0.77 & FW & 3.8     &  -4.03  &   67.3  & -3.73\\                                                                                    

& 25.02  & 7108.93 & -1.59 & FW & \ldots  &  \ldots &  24.7  & -3.53\\                                                                                    
& 25.02  & 7115.17 & -0.94 & FW & 4.0  &  -3.83   &  67.7  & -3.55\\
& 25.02  & 7116.99 & -0.91 & FW & 4.5  &  -3.82   &  68.4  & -3.57\\                                                                                    
& 25.02  & 7119.66 & -1.15 & FW & 5.5  &  -3.49   &  50.7  & -3.55\\

\noalign{\smallskip}                                                                
\label{C II}                                                     
C II & & & & \multicolumn{3}{c}{log C/N$_{T}$ = -3.78$\pm$0.04} & &  \ldots \\
\noalign{\smallskip}
                                                     
  &   &  &   &   &   &     &      &  \\
  &4  & 3918.97 & -0.53 & WF & 13.6 &-3.82    &   \ldots  & \ldots\\                   
  &6  & 4267.26 & +0.74 & WF & 28.8 & -3.74   &   \ldots  & \ldots \\                         
\noalign{\smallskip}                                                                
 \label{N I}                                                   
N I &&&&\multicolumn{3}{c}{log N/N$_{T}$ =-3.36$\pm$0.18}&& -3.61$\pm$0.12 \\ 
\noalign{\smallskip}                     
  
 &   &  &   &   &   &     &      &  \\ 

  

  &  6   & 4151.48 & -1.98 &WF &3.2   &-3.66   &   7.9  & -3.69\\                    
  & 9   & 4914.94 & -2.23 &WF & \ldots  &  \ldots   &   4.3  &    -3.44\\                    
  &     &  4935.12 & -1.89 &WF & \ldots  &  \ldots   &   5.2  &    -3.69\\                                                                                        
  \hline\noalign{\smallskip}                                                          
\end{tabular}                                                                                                                                                         
\\                                                                                                                                                        Note: gf value references follow: \\                                                
AT = \citet{2007A&A...461..767A};
B= \citet{1998ApJ...496.1051B}
BB= \citet{2007A&A...472L..43B};\\
BG = \citet{biemont1981}, \citet{biemont1989}; 
CB = \citet{1962etps.book.....C};CR = \citet{2007JPCRD..36.1287W}\\           

DS = \citet{1992A&A...255..457D} 
FW = \cite{fuhr2002} and \citet{1988JPCRD..17S....F};HL = \citet{hannaford1982}\\                                                                             

JK = \citet{1984PhRvA..30.2429J}
KG = \citet{kling2000};
KS = \citet{kling2001};KX = \citet{1995ASPC...78..205K}\\
LA = \citet{lanz1985};                                                               
LB = \citet{2001ApJ...556..452L} 
LD = \citet{lawler1989};LN = \citet{ljung2006}\\

LW = \citet{lawler2001};
MC = \citet{1975tsip.book.....M}
;NL = \citet{nil2006}\\

N4 = \citet{fuhr2006};
MF = \citet{1988JPCRD..17S....F} and \citet{1988atps.book.....M};RP = \citet{raassen1998}\\

PT = \citet{pickering2001};\citet{pickering2002};
PQ= \citet{2000PhyS...61..323P};SG = \citet{1969JQSRT...9...13S}\\

WF = \cite{wiese1996};
WM = \citet{wiese1980};                                                      
WS = \citet{wiese1969};\\

VALD,VALD2 data=  \citet{1995A&AS..112..525P}, \citet{1997BaltA...6..244R},  \citet{1999A&AS..138..119K}, \citet{2000BaltA...9..590K}           \\

-The lines marked with * are ignored in average calculations.\\
-This table is given electronically.

\end{flushleft}                                                                                                                                                       
\end{table*} 
\begin{table}[ht]
\center
\caption{The derived abundances and stellar parameters of HD\,80057 and HD\,80404 based on isochrones and evolutionary tracks of \citet{2013A&A...553A..24G} for $\Omega$ / $\Omega_{crit}$\,=0.65, 0.50 and our abundance analysis.\label{table:cno}}
\begin{tabular}{ l  c c}
\hline
 &HD80057 & HD80404\\
 \hline
 
 $\Omega$ & 0.60 & 0.50 \\
 N/C & 2.44$\pm$0.91 & 1.58$\pm$0.07 \\
predicted &2.39$\pm$0.05& 1.48$\pm$0.01 \\ 
 N/O& 0.43$\pm$0.17 & 0.50$\pm$0.30 \\
predicted & 0.52$\pm$0.01 & 0.49$\pm$0.12 \\ 
$\Sigma$\,$CNO$&  0.07$\pm$0.10 & -0.05$\pm$0.08 \\
  M/H& -0.15 $\pm$0.24 & -0.02 $\pm$ 0.20 \\
${\varv}$\,$_{crit}$ & 147$\pm$1 & 131$\pm$3\\
$vsini$ & 25$\pm$3 & 21$\pm$1\\
$R\,/\,R_\odot$			& 86$\pm$3 & 96$\pm$10\\
$M^{\rm spec}\,/\,M_\odot$ 		& 13$\pm$1&13$\pm$2\\
$M^{\rm ZAMS}\,/\,M_\odot $ 		& 13$\pm$1&12$\pm$2 \\
$M^{\rm evol}\,/\,M_\odot $		&13.00$\pm$0.30 & 11.80$\pm$0.20\\
log$\,L\,/\,L_\odot$			& 4.63$\pm$0.04 & 4.46$\pm$0.02\\[2mm]
   
  \hline
  
\end{tabular}
\end{table}

\appendix
\label{append}

\setcounter{table}{0} 
\renewcommand{\thetable}{A\arabic{table}}
\renewcommand{\thepage}{A\arabic{page}}
\begin{table*} 
\tiny                                                                     
\caption[ ]{Elemental Abundances of \astrobj{HD\,80057} and \astrobj{HD\,80404}\label{appendix}}                      

\begin{flushleft}                                                                   
    \begin{tabular}{cccccrcrc}  
    \hline                                                    
 & & & & &\multicolumn{2}{c}{\astrobj{HD\,80057}}&\multicolumn{2}{c}{\astrobj{HD\,80404}}\\         
Species&Multiplet& $\lambda$(\AA)& log gf&Ref.&W$_{\lambda}$(m{\AA})                
&log N/N$_{T}$ &  W$_{\lambda}$(m{\AA})& log N/N$_{T}$ \\                           
 \hline 
\noalign{\smallskip}

\label{CI}
\ion{C}{i} & & & & \multicolumn{3}{c}{log C/N$_{T}$ =-3.72$\pm$0.22} & &-3.65$\pm$0.14\\ 
                                                  
    \noalign{\smallskip}

& 6  & 4771.74 & -1.87 & FW & \ldots  &  \ldots &   74.4  & -3.47\\                       
&    & 4775.90 & -2.90 & FW & \ldots  &  \ldots &   37.7  & -3.47\\                       
& 13 & 5052.17 & -1.30 & CR & \ldots  &  \ldots &   69.9  & -3.90\\                       
&    & 4932.05 & -1.66 & FW & \ldots  &  \ldots &   47.2  & -3.84\\                       
& 14 & 4371.27 & -1.96 & FW & \ldots  &  \ldots &   31.8  & -3.75\\
\hline

\end{tabular}                                                                                                                                                           
\end{flushleft}                                                                                                                                                         
\end{table*}                                                                                                                                                           

                                                                                    
\setcounter{table}{0}                                                               
\begin{table*}                                                                                                                                                      
\tiny
\caption[ ]{- {\it continued}}                                                                                                             
\begin{flushleft}                                                                                                                                                      
\begin{tabular}{cccccrcrc}                                                          
                                                                                 
\hline                                                                              
                                                                                    
 & & & & &\multicolumn{2}{c}{\astrobj{HD\,80057}}&\multicolumn{2}{c}{\astrobj{HD\,80404}}\\         
Species&Multiplet& $\lambda$(\AA)& log gf&Ref.&W$_{\lambda}$(m{\AA})                
&log N/N$_{T}$ &  W$_{\lambda}$(m{\AA})& log N/N$_{T}$ \\                                        
                                                                                    
 \hline                                                                             
                 
\noalign{\smallskip}

\noalign{\smallskip}

\noalign{\smallskip}
\label{Si II}                                                       
Si II&{\it (continued)}&    &       &       &      &       &      & \\ 

 \noalign{\smallskip}  

&  5  & 5056.35 & -0.53 & WS &57.8  &-4.26  &   33.8  & -4.53\\                                                                                                                                                                                                                                                                                             &-    & 4075.45 & -1.40 & SG &25.0  &-4.66   &   6.8  & 5.05\\
&7.05 & 4621.42 & -0.54 & WS &13.1   &-4.77   &  2.3  & -4.82\\                       
&     & 4621.72 & -0.38 & WS &5.7  &-4.56   &   2.6  & -4.93\\                      
&7.15 & 4673.27 & -0.35 & KX &8.0  & -4.56   &   \ldots  & \ldots\\
\hline

\end{tabular}                                                                                                                                                           
\end{flushleft}                                                                                                                                                         
\end{table*}                                                                        
                                                                                    

\setcounter{table}{0}                                                                                                                                                   
\begin{table*}                                                                                                                                                          
\tiny  
\caption[ ]{- {\it continued}}                                                                                                                                          
\begin{flushleft}                                                                                                                                                      
\begin{tabular}{cccccrcrc}                                                          

\hline                                                                              
                                                                                    
  & & & & &\multicolumn{2}{c}{\astrobj{HD\,80057}}&\multicolumn{2}{c}{\astrobj{HD\,80404}}\\         
Species&Multiplet& $\lambda$(\AA)& log gf&Ref.&W$_{\lambda}$(m{\AA})                
&log N/N$_{T}$ &  W$_{\lambda}$(m{\AA})& log N/N$_{T}$ \\                                      
                                                                                    
 \hline                                                                             
                                                     
 \noalign{\smallskip}                                                               
                                                               

 \label{Ti I}                                                      
Ti I&&&&\multicolumn{3}{c}{log Ti/N$_{T}$ = \ldots}&& -7.53$\pm$0.21\\
&  6  & 4656.46 & -1.35 &WS  & \ldots  &  \ldots   &   6.2  &      -6.63\\  
&     & 4667.58 & -1.19 &WS  & \ldots  &  \ldots   &   3.5  &      -7.03\\ 
&     & 4681.90 & -1.07 &WS  & \ldots  &  \ldots   &   5.2  &      -6.96\\ 
& 11  & 3982.48 & -1.27 &WS  & \ldots  &  \ldots   &   10.7  &      -6.40\\ 
& 12  & 3981.76 & -0.35 &WS  & \ldots  &  \ldots   &   56.4  &      -6.37\\
\hline 
\end{tabular}                                                                                                                                                           
\end{flushleft}                                                                                                                                                         
\end{table*}                                                                        
                                                                  
                                                                                    
\setcounter{table}{0}                                                                                                                                                  
\begin{table*}                                                                                                                                                          
\tiny  
\caption[ ]{- {\it continued}}                                                                                                                                          
\begin{flushleft}                                                                                                                                                      
\begin{tabular}{cccccrcrc}                                                          

\hline                                                                              
                                                                                    
  & & & & &\multicolumn{2}{c}{\astrobj{HD\,80057}}&\multicolumn{2}{c}{\astrobj{HD\,80404}}\\         
Species&Multiplet& $\lambda$(\AA)& log gf&Ref.&W$_{\lambda}$(m{\AA})                
&log N/N$_{T}$ &  W$_{\lambda}$(m{\AA})& log N/N$_{T}$ \\                                      
                                                                                    
 \hline                                                                             

 Ti II &{\it (continued)}&    &      &        &      &       &      & \\

 &60  & 4568.31 & -2.94 & PT &\ldots& \ldots &   42.8  & -7.08\\ 
 &    & 4580.45 & -2.94 & PT &3.2   &-7.11   &   33.9  & -7.21\\
 &    & 4600.26 & -3.54 & VALD2 &\ldots& \ldots &   \ldots & \ldots\\
 &61  & 4391.03 & -2.28 & PT &5.9   &-7.49   &100.2   &-6.89\\                      
 &    & 4411.93 & -2.52 & PT &3.2   &-7.52   & 58.0.  &-7.27\\
\end{tabular}                                                                                                                                                   
\end{flushleft}                                                                                                                                      
\end{table*}                                                                       
                                                                                    
\setcounter{table}{0}                                                                                                                                                   
\begin{table*}                                                                                                                                                         
\tiny  
\caption[ ]{- {\it continued}}                                                                                                                                          
\begin{flushleft}                                                                                                                                                       
\begin{tabular}{cccccrcrc}                                                          

\hline                                                                              
                                                                                    
  & & & & &\multicolumn{2}{c}{\astrobj{HD\,80057}}&\multicolumn{2}{c}{\astrobj{HD\,80404}}\\         
Species&Multiplet& $\lambda$(\AA)& log gf&Ref.&W$_{\lambda}$(m{\AA})                
&log N/N$_{T}$ &  W$_{\lambda}$(m{\AA})& log N/N$_{T}$ \\                              
                                                                                    
\hline

\noalign{\smallskip}                                                                                                                                                   
\label{Cr I}
Cr I &{\it (continued)} & & & \multicolumn{3}{c}{log Cr/N$_{T}$ = \ldots} & & -6.40$\pm$0.17\\          
\noalign{\smallskip}                                                                                                                                                   
& 18    & 5298.27 & -1.15 & MF & \ldots  &  \ldots   &  19.7  &     -6.38\\                                                                                        
&     & 5300.74 & -2.12 & MF & \ldots  &  \ldots   &   4.9  &     -6.07\\                                                                                        
&     & 5345.80 & -0.98 & MF & \ldots  &  \ldots   &  31.1  &     -6.29\\                                                                                                                                                                                
&     & 5348.31 & -1.29 & MF & \ldots  &  \ldots   &  16.8  &     -6.30 \\                                                                                        
&     & 5409.77 & -0.72 & MF & \ldots  &  \ldots   &  38.8&     -6.40\\
\end{tabular}                                                                                                                                                          
\end{flushleft}                                                                                                                                                        
\end{table*}                                                                        
 
                                                                                    
\setcounter{table}{0}                                                               
                                                                                    
\begin{table*}                                                                      
\tiny
\caption[ ]{- {\it continued}}                                                      
                                                                                    
\begin{flushleft}                                                                   
                                                                                    
\begin{tabular}{cccccrcrc}                                                          
\hline                                                                              
                                                                                    
   & & & & &\multicolumn{2}{c}{\astrobj{HD\,80057}}&\multicolumn{2}{c}{\astrobj{HD\,80404}}\\         
Species&Multiplet& $\lambda$(\AA)& log gf&Ref.&W$_{\lambda}$(m{\AA})                
&log N/N$_{T}$ &  W$_{\lambda}$(m{\AA})& log N/N$_{T}$ \\                               
                                                                                    
 \hline 
\noalign{\smallskip}                                                                                                                                                   
Cr II&{\it (continued)}&&&\multicolumn{3}{c}{log Cr/N$_{T}$ = -6.46$\pm$0.20}&&-6.56$\pm$0.26\\      
\noalign{\smallskip}
\label{Cr II}     

&44    & 4592.04 & -1.42 & NL   &51.2    &-6.64   &   97.7    &     -6.49\\                      
&     & 4616.63 & -1.58 & VALD &45.3    &-6.56   &   91.7    &     -6.43\\                      
&     & 4618.80 & -1.00 & NL   &89.2    &-6.68   &   120.0   &     -6.51\\                      
&     & 4634.07 & -1.24 & MF   &75.5    &-6.57   &   111.0   &     -6.44\\                    
&177  & 4697.60 & -1.88 & MF   & \ldots &\ldots  &    13.6   &     -6.34\\
\end{tabular}                                                                                                                                                         
\end{flushleft}                                                                                                                                                       
\end{table*}


\setcounter{table}{0}                                                                                                                                                   
\begin{table*}                                                                                                                                                        
\tiny  
\caption[ ]{- {\it continued}}                                                                                                                                         
\begin{flushleft}                                                                                                                                                      
\begin{tabular}{cccccrcrc}                                                          

\hline                                                                                                                                                                 
  & & & & &\multicolumn{2}{c}{\astrobj{HD\,80057}}&\multicolumn{2}{c}{\astrobj{HD\,80404}}\\         
Species&Multiplet& $\lambda$(\AA)& log gf&Ref.&W$_{\lambda}$(m{\AA})                
&log N/N$_{T}$ &  W$_{\lambda}$(m{\AA})& log N/N$_{T}$ \\                                      
 \hline

Fe I &{\it (continued)}&    &      &       &      &       &      & \\                                                                                       
\noalign{\smallskip}
  &    &         &        &   &        &      &&\\                  
 & 16 & 5041.07& -3.09 & VALD & \ldots  &  \ldots   &  69.6  &    -4.70\\ 
 &    & 5051.63 & -2.80 & N4 & \ldots  &  \ldots   &   44.1  &    -4.44\\                        
 &    & 5079.73 & -3.22 & N4 & \ldots  &  \ldots   &   24.5  &    -4.31\\                                                                                             
 & 20 & 3840.44 & -0.51 & N4 &24.7  &-4.94   &  132.7  & -4.87\\
\end{tabular}                                                                                                                                                          
\end{flushleft}                                                                                                                                                         
\end{table*}                                                                        
                                                                                    
\setcounter{table}{0}                                                               
                                                                                    
\begin{table*}                                                                      
\tiny                                                                                
\caption[ ]{- {\it continued}}                                                      
                                                                                    
\begin{flushleft}                                                                   
                                                                                    
\begin{tabular}{cccccrcrc}                                                          
\hline                                                                              
                                                                                    
 & & & & &\multicolumn{2}{c}{\astrobj{HD\,80057}}&\multicolumn{2}{c}{\astrobj{HD\,80404}}\\         
Species&Multiplet& $\lambda$(\AA)& log gf&Ref.&W$_{\lambda}$(m{\AA})                
&log N/N$_{T}$ &  W$_{\lambda}$(m{\AA})& log N/N$_{T}$ \\                                     
                                                                                    
 \hline                                                                             
 
 \noalign{\smallskip}                                                                                                                                                   
      
\label{Fe I}                                                                                                                                                                       
Fe I &{\it (continued)}&    &      &       &      &       &      & \\                                                                             
                       
  \noalign{\smallskip}               
               
 &522 & 4199.10 & +0.16 & N4 &9.6   &-4.84  &   94.5  &     -4.83\\             
 &528 & 3843.26 & -0.24 & N4 &4.8  & -4.72   &   71.0  &    -4.819\\                      
 &664 & 3846.80 & -0.02 & N4 &4.5   &-4.85   &   71.7  &     -4.88\\                        
 &    & 3906.75 & -0.95 & N4 & \ldots  &  \ldots   &   30.5  & -4.62\\   
  &686  &5615.64&   0.05 & N4 &\ldots  &\ldots &103.0     &    -4.488 \\
\end{tabular}                                                                                                                                                         
\end{flushleft}                                                                                                                                                         
\end{table*}

\setcounter{table}{0}                                                               
                                                                                    
\begin{table*}                                                                      
\tiny  
\caption[ ]{- {\it continued}}                                                      
                                                                                    
\begin{flushleft}                                                                   
                                                                                    
\begin{tabular}{cccccrcrc}                                                          
\hline                                                                              
                                                                                    
  & & & & &\multicolumn{2}{c}{\astrobj{HD\,80057}}&\multicolumn{2}{c}{\astrobj{HD\,80404}}\\         
Species&Multiplet& $\lambda$(\AA)& log gf&Ref.&W$_{\lambda}$(m{\AA})                
&log N/N$_{T}$ &  W$_{\lambda}$(m{\AA})& log N/N$_{T}$ \\                                      
                                                                                    
 \hline

 \label{Fe II}

Fe II &{\it (continued)}&    &       &       &      &       &      & \\

 & 32  & 4439.13 & -5.50 & VALD & \ldots  &  \ldots   &   6.2  &    -4.50\\ 
 & 36 & 4993.35 & -3.68 & N4 &38.5  &-4.78   &   \ldots  & \ldots\\
 & 37 & 4472.92 & -3.53 & N4 &58.3  &-4.66   &   82.1  &    -4.75\\                     
 &    & 4489.18 & -2.97 & N4 &122.8 &-4.58   &   \ldots  & \ldots\\ 
 &    & 4491.40 & -2.64 & N4 & 152.0  &  -4.58   &   \ldots  & \ldots\\
\end{tabular}                                                                                                                                                         
\end{flushleft}                                                                                                                                                       
\end{table*}                                                                        
\setcounter{table}{0}                                                               

\begin{table*}                                                                      
\tiny
\caption[ ]{- {\it continued}}                                                      
                                                                                    
\begin{flushleft}                                                                   
                                                                                    
\begin{tabular}{cccccrcrc}                                                          
\hline                                                                              
                                                                                    
   & & & & &\multicolumn{2}{c}{\astrobj{HD\,80057}}&\multicolumn{2}{c}{\astrobj{HD\,80404}}\\         
Species&Multiplet& $\lambda$(\AA)& log gf&Ref.&W$_{\lambda}$(m{\AA})                
&log N/N$_{T}$ &  W$_{\lambda}$(m{\AA})& log N/N$_{T}$ \\                                
                                                                                    
 \hline

 \label{Fe II-D-G-J}

Fe II &{\it (continued)}&    &       &       &      &       &      & \\                   
 & J   & 4451.55 & -1.91 & VALD & \ldots  &  \ldots   &     57.1 &    -4.42\\                      
 &    & 4455.27 & -2.00 & VALD & \ldots  &  \ldots   &     43.4 &    -4.49\\    
 &   & 4480.68 & -2.56 & VALD & \ldots  &  \ldots   &  19.6  &   -4.41\\                                                                                   
 &    & 4579.53 & -2.34 & VALD & 22.5  &  -4.45   &   \ldots  & \ldots\\                      
 &    & 4640.84 & -1.74 & VALD & 6.7  &  -4.82   &   \ldots  & \ldots\\
\end{tabular}                                                                                                                                                         
\end{flushleft}                                                                                                                                                       
\end{table*}  

\clearpage
\setcounter{table}{0}                                                               
                                                                                    
\begin{table*}                                                                      
\tiny
\caption[ ]{- {\it continued}}                                                      
                                                                                    
\begin{flushleft}                                                                   
                                                                                    
\begin{tabular}{cccccrcrc}                                                          
\hline                                                                              
                                                                                    
   & & & & &\multicolumn{2}{c}{\astrobj{HD\,80057}}&\multicolumn{2}{c}{\astrobj{HD\,80404}}\\         
Species&Multiplet& $\lambda$(\AA)& log gf&Ref.&W$_{\lambda}$(m{\AA})                
&log N/N$_{T}$ &  W$_{\lambda}$(m{\AA})& log N/N$_{T}$ \\                                
                                                                                    
 \hline 
 \noalign{\smallskip}    
Ni I &{\it (continued)}&    &       &       &      &       &      & \\ 
 &117 & 3908.91& -0.57 & KX &\ldots & \ldots  &  10.1  & -5.73 \\
 &130 & 4855.40&  0.00 & FW &\ldots & \ldots  &  25.7  & -5.95 \\
 &131 & 4829.01& -0.33 & FW &\ldots & \ldots  &  21.6  & -5.71 \\
 &132 & 4752.43& -0.70 & FW &\ldots & \ldots  &   9.1  & -5.68 \\
 &    & 4913.96& -0.63 & FW &\ldots & \ldots  &   9.6  & -5.65\\
\end{tabular}                                                                                                                                                         
\end{flushleft}                                                                                                                                                       
\end{table*}


                                                                                    
\setcounter{table}{0}                                                               
                                                                                    
\begin{table*}                                                                      
\tiny                                                                                   
\caption[ ]{- {\it continued}}                                                      
                                                                                    
\begin{flushleft}                                                                   
                                                                                    
\begin{tabular}{cccccrcrc}                                                          
\hline                                                                              
                                                                                    
  & & & & &\multicolumn{2}{c}{\astrobj{HD\,80057}}&\multicolumn{2}{c}{\astrobj{HD\,80404}}\\         
Species&Multiplet& $\lambda$(\AA)& log gf&Ref.&W$_{\lambda}$(m{\AA})                
&log N/N$_{T}$ &  W$_{\lambda}$(m{\AA})& log N/N$_{T}$ \\                               
                                                                                    
 \hline

Eu II&&&&\multicolumn{3}{c}{log Eu/N$_{T}$ = \ldots} & &-11.74$\pm$0.13\\                
\label{Eu II} 

 &  1 & 3819.67 & +0.51 & LW & \ldots  &  \ldots   &   24.7  & -11.82\\
 &    & 4129.70 & +0.22 & LW & \ldots  &  \ldots   &   18.7  & -11.72\\
 &    & 4205.05 & +0.21 & LW & \ldots  &  \ldots   &   26.3  & -11.54\\
 &  5 & 3907.10 & +0.17 & LW & \ldots  &  \ldots   &   8.2   & -11.88\\

\hline\noalign{\smallskip}                                                          
\end{tabular}                                                                                                                                                         
\\                                                                                                                                                        Note: gf value references follow: \\                                                
AT = \citet{2007A&A...461..767A};
B= \citet{1998ApJ...496.1051B}
BB= \citet{2007A&A...472L..43B};\\
BG = \citet{biemont1981}, \citet{biemont1989}; 
CB = \citet{1962etps.book.....C};CR = \citet{2007JPCRD..36.1287W}\\           

DS = \citet{1992A&A...255..457D} 
FW = \cite{fuhr2002} and \citet{1988JPCRD..17S....F};HL = \citet{hannaford1982}\\                                                                             

JK = \citet{1984PhRvA..30.2429J}
KG = \citet{kling2000};
KS = \citet{kling2001};KX = \citet{1995ASPC...78..205K}\\
LA = \citet{lanz1985};                                                               
LB = \citet{2001ApJ...556..452L} 
LD = \citet{lawler1989};LN = \citet{ljung2006}\\

LW = \citet{lawler2001};
MC = \citet{1975tsip.book.....M}
NL = \citet{nil2006}\\

N4 = \citet{fuhr2006};
MF = \citet{1988JPCRD..17S....F} and \citet{1988atps.book.....M};RP = \citet{raassen1998}\\

PT = \citet{pickering2001};\citet{pickering2002};
PQ= \citet{2000PhyS...61..323P};SG = \citet{1969JQSRT...9...13S}\\

WF = \cite{wiese1996};
WM = \citet{wiese1980};                                                      
WS = \citet{wiese1969};\\

VALD,VALD2 data=  \citet{1995A&AS..112..525P}, \citet{1997BaltA...6..244R}, \citet{1999A&AS..138..119K}\\ 
 \citet{2000BaltA...9..590K}           \\

\end{flushleft}                                                                                                                                                       
\end{table*}                                                                        


                         
\end{document}